\documentclass[preprints,article,accept,oneauthor,pdftex]{myCls} 

\firstpage{1} 
\pubvolume{1}
\issuenum{1}
\articlenumber{0}
\pubyear{2022}
\copyrightyear{2022}
\hreflink{https://doi.org/} % If needed use \linebreak

\newcommand{\Figname}{Figure~}
\newcommand{\Figsname}{Figures~}

\newcommand{\secname}{Section~}
\newcommand{\secsname}{Sections~}
\newcommand{\gl}{Equation}
\newcommand{\gls}{Equations}
\newcommand{\Gl}{Equation}
\newcommand{\Gls}{Equations}
\usepackage[version=4]{mhchem} % Formula subscripts using \ce{}
\usepackage{threeparttable}
\DeclareGraphicsExtensions{.jpg, .eps, .pdf, .png}

\graphicspath{{/media/ioan/Elements/gt/antioxidant/data/}{.}}
\Title{% Comprehensive Quantum Chemical Characterization of
  Why Ortho- and Para-Hydroxy Metabolites Can Scavenge Free Radicals that the Parent Atorvastatin
  Cannot? Important Pharmacologic Insight from Quantum Chemistry
  }
\TitleCitation{ Why Ortho- and Para-Hydroxy Metabolites Can Scavenge Free Radicals that the Parent Atorvastatin
  Cannot? Important Pharmacologic Insight from Quantum Chemistry}

\Author{Ioan B\^aldea % \orcidA{}
}
\AuthorNames{Ioan B\^aldea}

\AuthorCitation{B\^aldea, I.}
\address[1]{%
  Theoretical Chemistry, Heidelberg University, Im Neuenheimer Feld 229, D-69120 Heidelberg, Germany; ioan.baldea@pci.uni-heidelberg.de
}
\abstract{The pharmaceutical success of atorvastatin (ATV),
  a widely employed drug against the ``bad'' cholesterol (LDL) and cardiovascular diseases,
  traces back to its ability to scavenge free radicals.
  Unfortunately, information on its antioxidant properties is missing or unreliable.
  Here, we report detailed quantum chemical results for ATV and its   
  ortho- and para-hydroxy metabolites (o-ATV, p-ATV) in the methanolic phase.
  They comprise global reactivity indices,
  bond order indices, and spin densities as well as all relevant enthalpies of reaction
  (bond dissociation BDE, ionization IP and electron attachment EA,
  proton detachment PDE and proton affinity PA, and electron transfer ETE).
  With these properties in hand, we can provide the first theoretical explanation of the experimental finding that, due to their free radical
  scavenging activity, ATV hydroxy metabolites rather than the parent ATV, have substantial inhibitory effect on LDL and the like. 
  Surprisingly (because it is contrary to the most cases currently known),
  we unambiguously found that HAT (direct hydrogen atom transfer) rather than SPLET (sequential proton loss electron transfer) or
  SET-PT (stepwise electron transfer proton transfer) 
  is the thermodynamically preferred pathway
  by which o-ATV and p-ATV in methanolic phase can scavenge 
  DPPH$^\bullet$ %mdpi:please confirm if the \bullet is correct? check it in the whole article. Author: It's OK.
  (1,1-diphenyl-2-picrylhydrazyl) radicals.
  From a quantum chemical perspective, the ATV's species investigated are surprising because of the nontrivial correlations between
  bond dissociation energies,
  bond lengths, bond order indices and pertaining stretching frequencies,
  which do not fit the framework of naive chemical intuition.
}


\begin{document}

\section{Introduction}
\label{sec:intro}
The highly radical scavenging active cholesterol-lowering drug atorvastatin (ATV) \cite{Roth:02}
is an outstanding % incredible
success sale story \cite{atv-statista}.
It was patented in 1985 and approved by the Food and Drug Administration (FDA) in 1996 for medical use.
Sold under the name of Lipitor by the world's leading pharmaceutical company Pfizer,
it received record high revenues of about 12.8 billion US dollars in 2006,
still generated ten billion US dollars in the year of patent loss (2011) and nearly two billion US dollars in 2019.
ATV, one of the most prescribed drugs in the US today, is mainly employed to prevent high risk for
developing cardiovascular diseases and as treatment for abnormal lipid levels (dyslipidemia).
ATV's inhibition of the HMG-CoA (3 hydroxy-3-methylglutaryl coenzyme A) reductase is
plausibly related to the high radical scavenging potency against lipoprotein oxidation.

ATV made the object of several theoretical investigations in the past \cite{Alnajjar:21,Hoffmann:08}.
Still, the antioxidant properties of ATV
were only recently investigated from the quantum chemical perspective  \cite{Duque:22}.
Unfortunately, as we drew attention recently \cite{baldea_2022_chemrxiv}, the only quantum chemical
attempt of which we are aware \cite{Duque:22} is plagued by severe flaws \cite{baldea_2022_chemrxiv}
(e.g., ``prediction'' of enormous, totally unrealistic O-H bond dissociation energies of $\sim 400\,\mbox{kcal/mol} > 17\,$eV),
and this makes mandatory the effort (undertaken in the present paper) of properly reconsidering the antioxidant capacity of ATV and
its ortho- and para-hydroxy metabolites in methanol.
For the notoriously poor soluble ATV, this solvent is of special interest.
ATV is freely soluble in methanol. In addition, antioxidant assays are mostly done in methanolic environment \cite{Portes:07,Duque:22}.
Along with quantities traditionally related to the antioxidant activity, 
the present study will also reports on the ATV global chemical reactivity indices, relevant bond data as well as spin densities
of radical species generated by H-atom abstraction from ATV and related ortho- and para-hydroxylated derivatives (o-ATV,
p-ATV, respectively). 

Theoretical understanding of the differences between ATV and its ortho- and para-hydroxy metabolites, which is missing to date,
is of paramount practical importance. A twenty four years old experimental study reported that atorvastatin ortho- and para-hydroxy metabolites
(o-ATV and p-ATV, respectively) protect, e.g., LDL from oxidation, while the parent ATV does not
\cite{Aviram:98}. Importantly for the results we are going to present in \secname\ref{sec:practice},
the free radical scavenging activity of o-ATV and p-ATV was analyzed by the ubiquitous
1,1 diphenyl-2 picryl-hydrazyl (DPPH$^\bullet$) assay in ref.~\cite{Aviram:98}.
Our study is able to provide the first theoretical explanation of this experimental finding.
\section{Computational Details}
\label{sec:methods}
The results reported below were obtained from quantum chemical calculations
wherein all necessary steps (geometry optimizations, frequency calculations, and electronic energies)
where conducted at the same DFT level of theory by running GAUSSIAN 16 \cite{g16}
on the bwHPC platform \cite{bwHPC}. In all cases investigated, we convinced ourselves that all frequencies are real.
In all calculations we used  6-31+G(d,p) basis sets \cite{Petersson:88,Petersson:91} and,
unless otherwise specified
(see \secname\ref{sec:eta} and \ref{sec:bde}),
the hybrid B3LYP exchange correlation functional \cite{Parr:88,Becke:88,Becke:93a,Frisch:94}.

For comparative purposes, we also present results obtained by using the PBE0 \cite{Adamo:99} 
functional and Truhlar's M062x \cite{Truhlar:06,Truhlar:08}
(see \secname\ref{sec:eta} and \ref{sec:bde}).
Computations for open shell species were carried out using unrestricted spin methods (e.g., UB3LYP and UPBE0).
In most radicals, employing the more computationally demanding quadratic convergence SCF methods was unavoidable.
We convinced ourselves that spin contamination is not a severe issue. In all these calculations, we invariably found a value 
$\left\langle S^2\right\rangle = 0.7501$ for the total spin after annihilation of the first spin contaminant,
versus the exact value $\left\langle S^2\right\rangle = 3/4 $.

Still, to better check this aspect, for ATV's cation and anion as well as for the ATV1H
and ATV4H radicals (see \secname\ref{sec:geometry} for the meaning of these acronyms) we also undertook the rare
numerical effort (enormous for molecules with almost 80 atoms)
of performing \emph{full} restricted open shell (ROB3LYP) calculations;
that is, not only single point calculations for electronic energy but also geometry optimization and (numerical) vibrational
frequency calculations, and all these in solvent. Differences between UB3LYP and ROB3LYP were reasonably small
(see \secname\ref{sec:eta} and \ref{sec:bde}),
but they should make it clear that claims
(so often formulated in the literature on antioxidation) of chemical accuracy ($\sim$1\,kcal/mol) at the 
B3LYP/6-31+G(d,p) are totally out of place.
From experience with much smaller molecules and much simpler chemical structures (e.g.,~ref.~\cite{Baldea:2022e})
we had to learn that achieving this accuracy for bond dissociation enthalpies and proton affinity (BDE and PA, quantities entering
the discussion that follows) is often illusory even for extremely computationally demanding
state-of-the-art compound model chemistries (CBS-QB3, CBS-APNO, G4, W1BD); see, e.g., Figure 10 of ref.~\cite{Baldea:2022e}.
   {DFT-calculations done by us and by others \cite{Kaiser:10} revealed that, e.g., errors in ionization potential can amount up to
0.7\,eV (16\,kcal/mol) even when employing the functional B3LYP and the largest Pople basis set
6-311++G(3df,3pd).}

Unless otherwise specified, the solvent (methanol) was accounted for
within the polarized continuum model (PCM) \cite{Tomasi:05} using the integral equation formalism (IEF) \cite{Cances:97}.
Although this is the ``gold standard'' for modeling solvents in the literature on free radical scavenging, one should be aware that 
this framework ignores specific solvation effects (hydrogen bonds). Because they may play an important role, e.g., in proton transfer reactions, 
theoretical estimates of PA may not be sufficiently accurate. While this makes comparison with experiment problematic, it should be a less critical issue
when comparing among themselves PA values of various antioxidants in a given solvent (e.g., methanol). To better emphasize 
why we believe that solvent effects in the context of antioxidants deserve a more careful consideration, along with IEFPCM-based results,
we also present results obtained in Truhlar's SMD solvation model \cite{Truhlar:08a, Truhlar:08b,Truhlar:09}.

GABEDIT \cite{gabedit} was used to generate molecular geometries and spatial distributions 
from the GAUSSIAN output (*.log) files.
To compute Wiberg bond order indices, we used the package NBO 6.0~\cite{NBO:6.0} interfaced with GAUSSIAN 16.
The reason why we use Wiberg bond order indices~\cite{Wiberg:68} rather than the heavily advertised Mayer bond order indices \cite{Mayer:07} was explained elsewhere~\cite{baldea_2022b}.
All thermodynamic properties were calculated at $T = 298.15$\,K.
\begin{figure}[H]

{
\includegraphics[width=5.5cm]{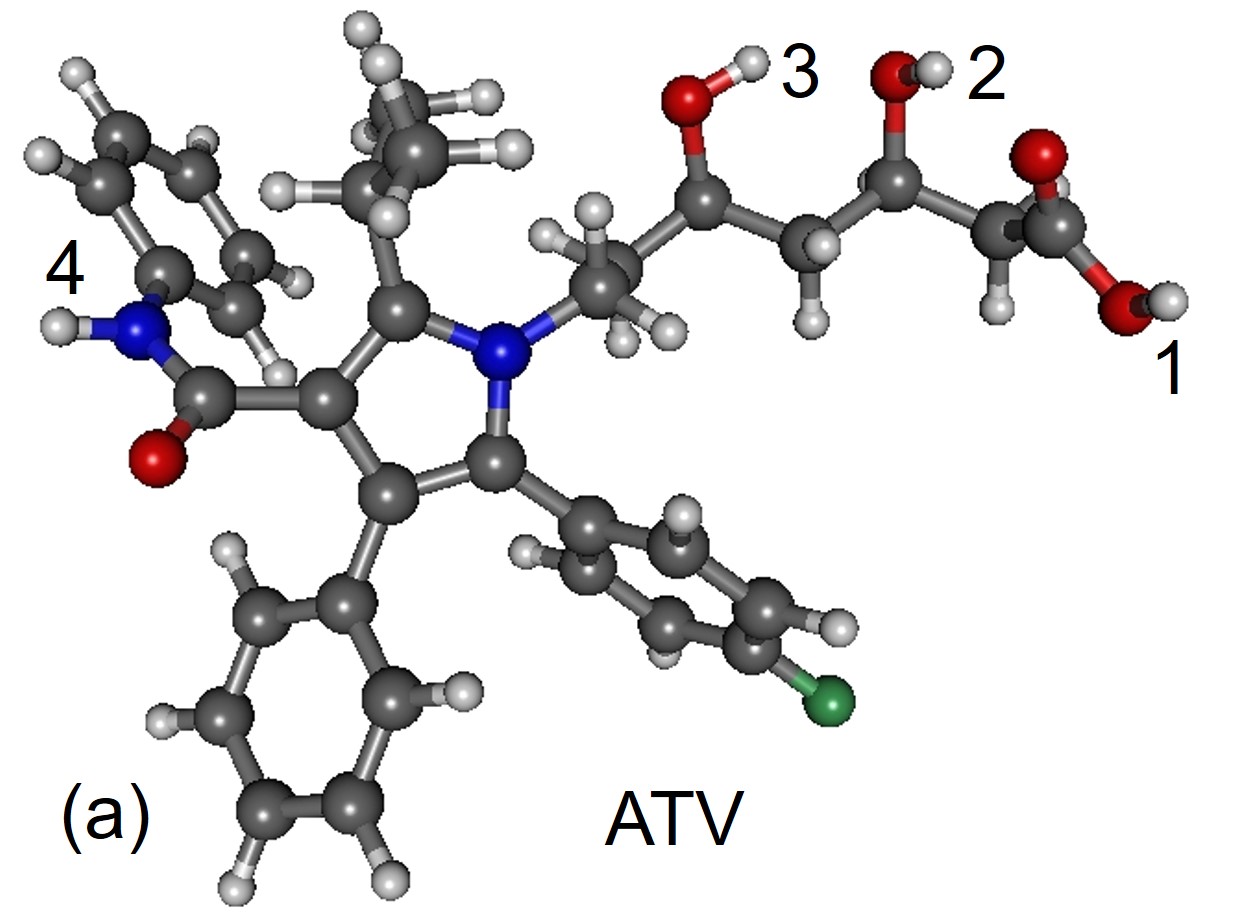}
\includegraphics[width=5.5cm]{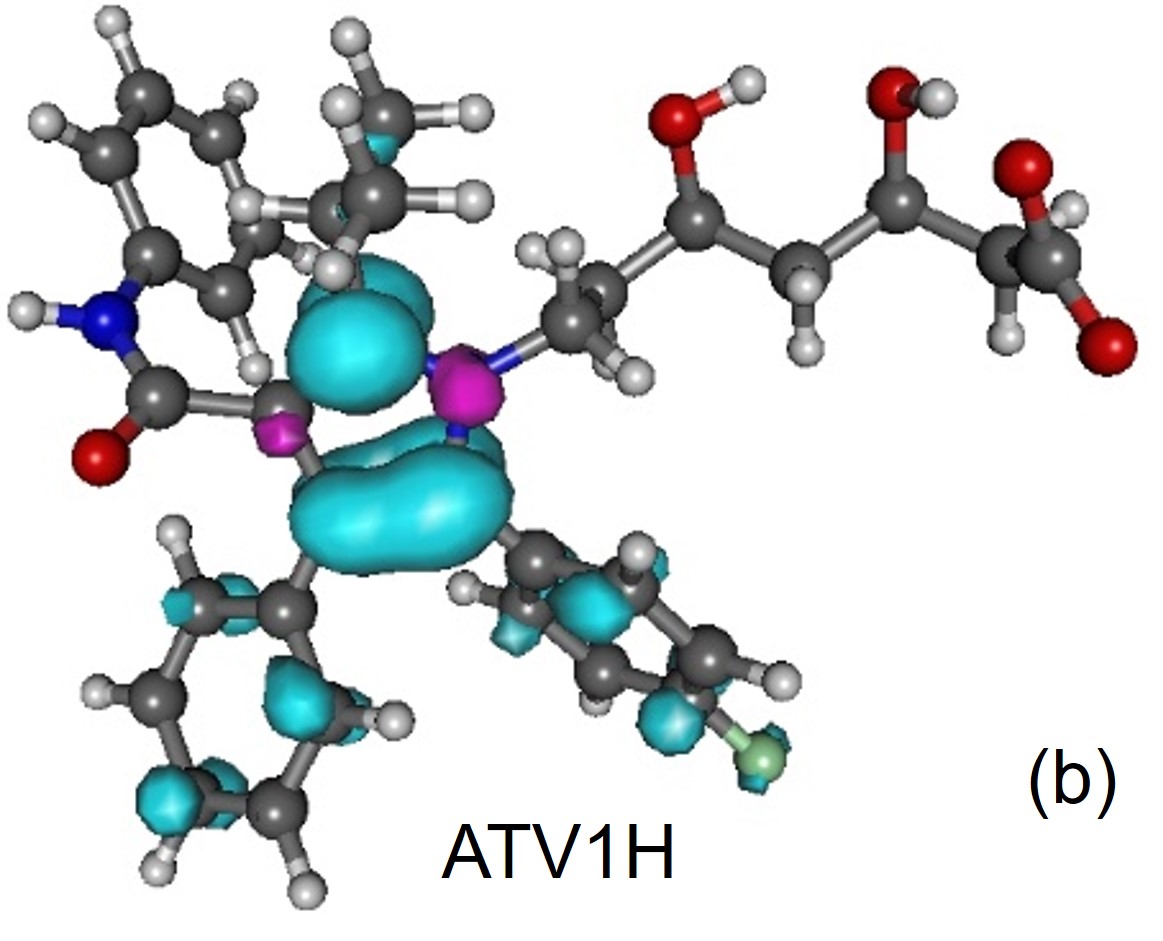}
}\\
{
\includegraphics[width=5.5cm]{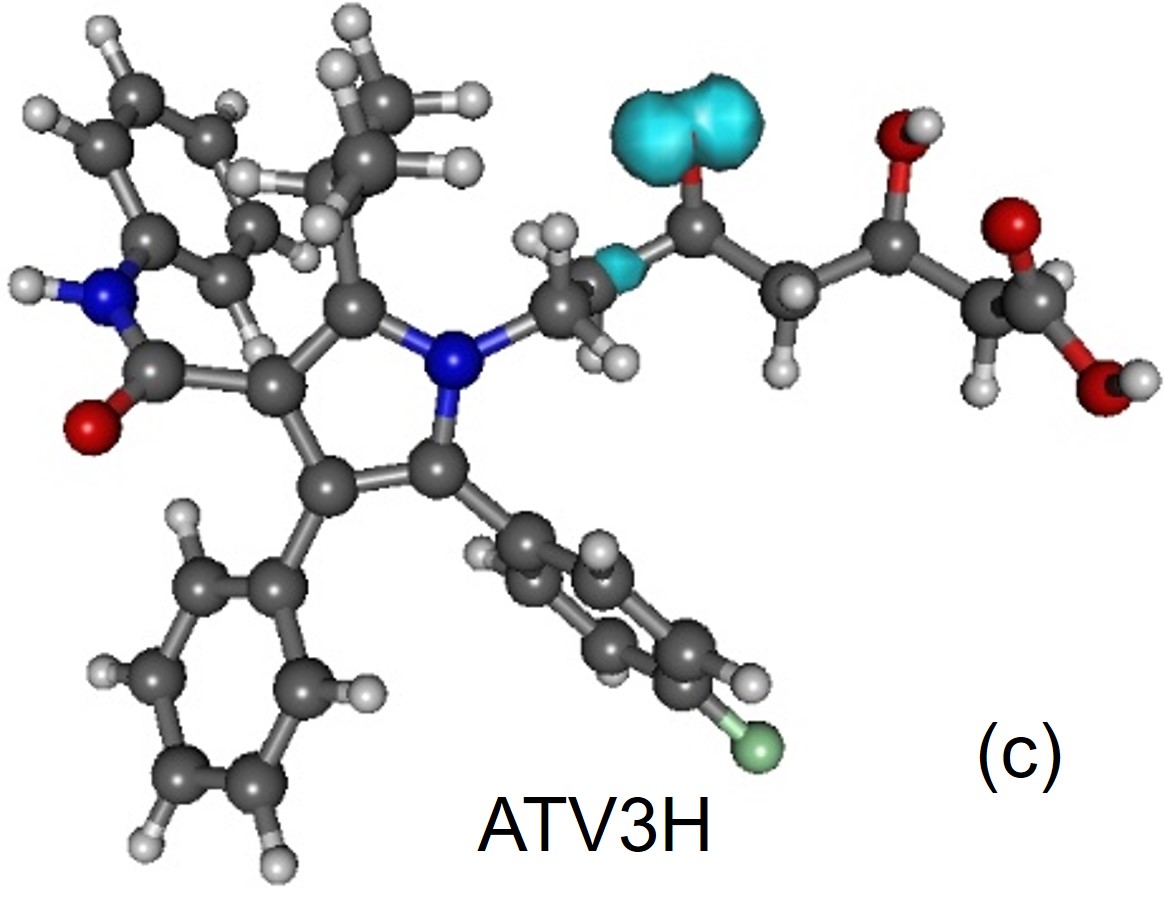}
\includegraphics[width=6.3cm]{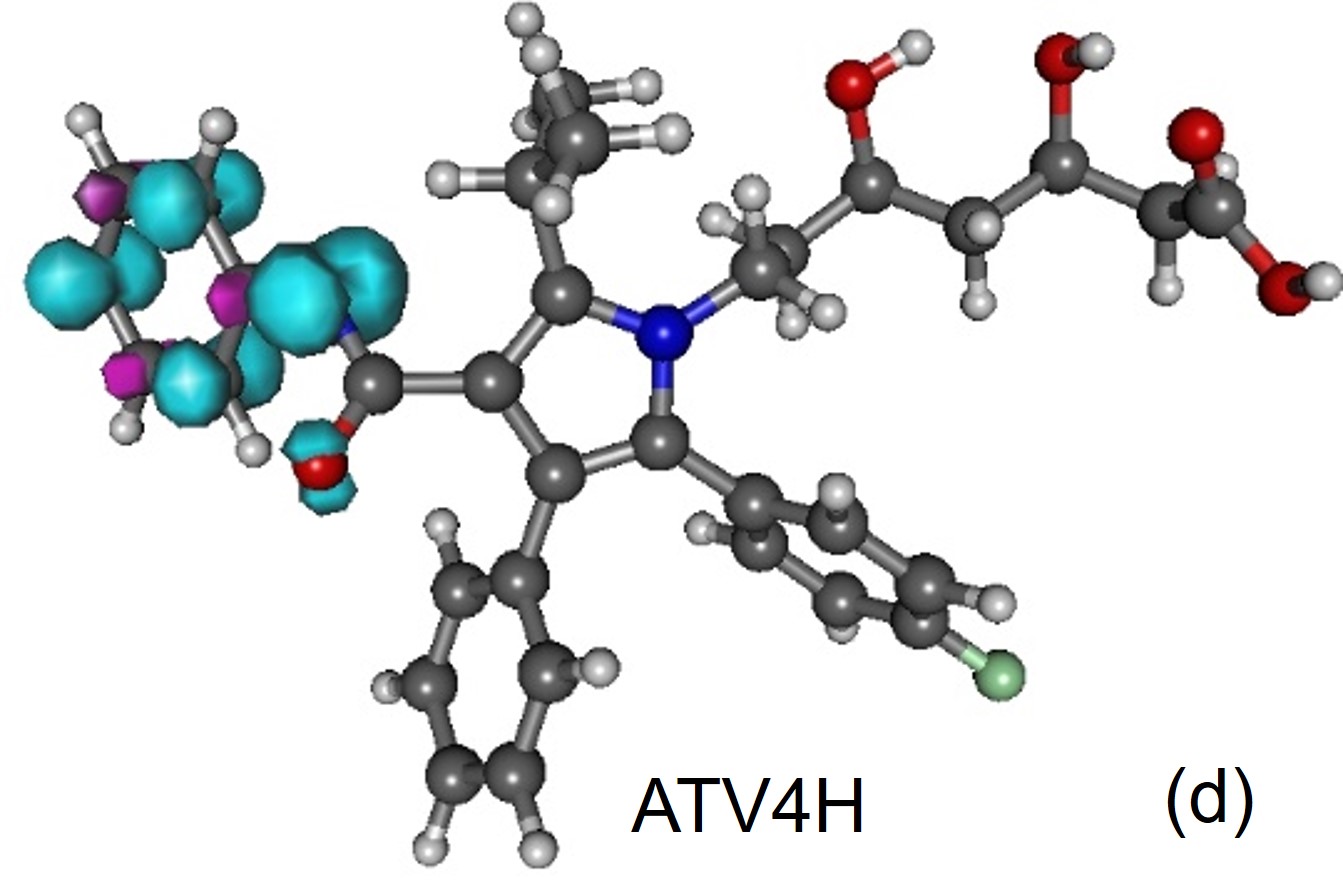}
}

\caption{(\textbf{a}) Optimized ATV geometry. Spin densities of neutral radicals generated from it by H-atom abstraction at positions
  indicated in the inset: (\textbf{b}) ATV1H, (\textbf{c}) ATV3H, and (\textbf{d}) ATV4H.
  Figure generated using GABEDIT \cite{gabedit}.\label{fig:atv}}
\end{figure}
\vspace{-6pt}

\begin{figure}[H]
{
\includegraphics[width=5.9cm]{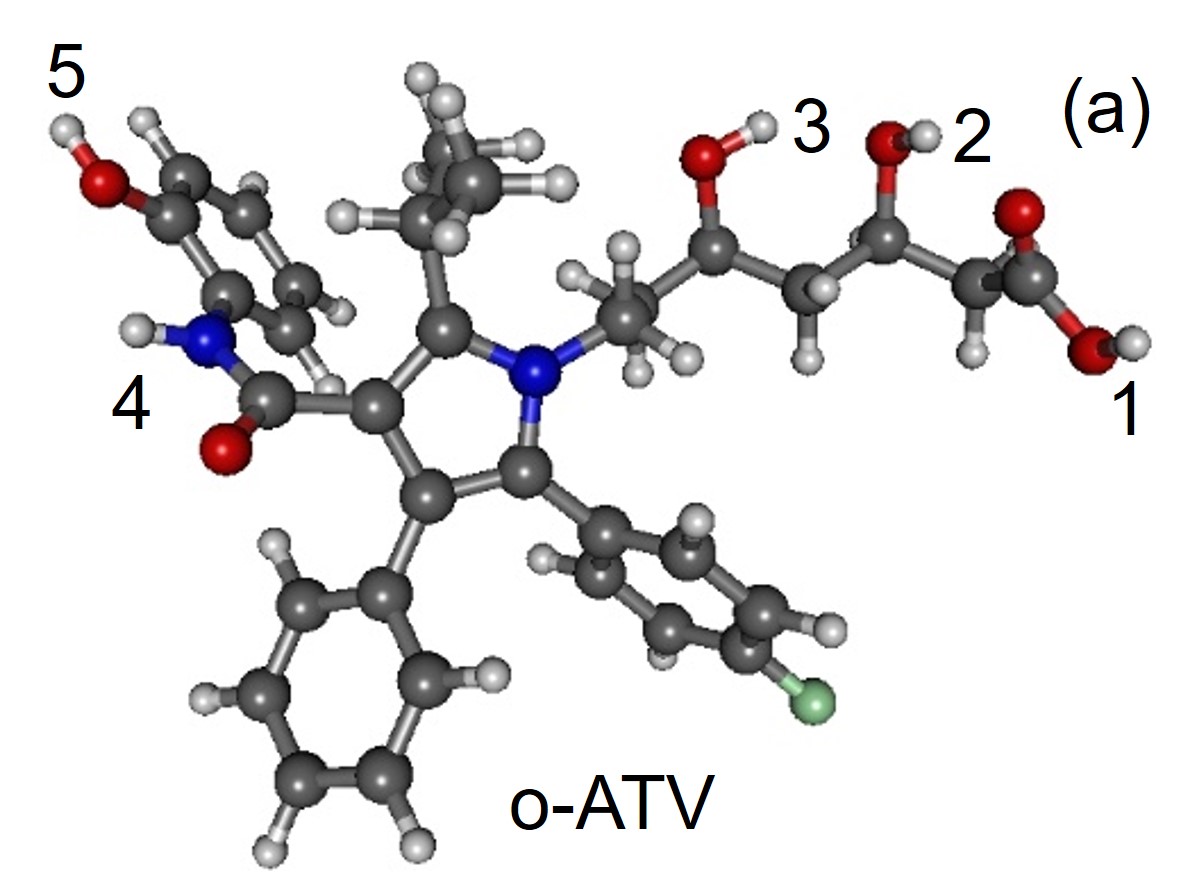}
\includegraphics[width=5.9cm]{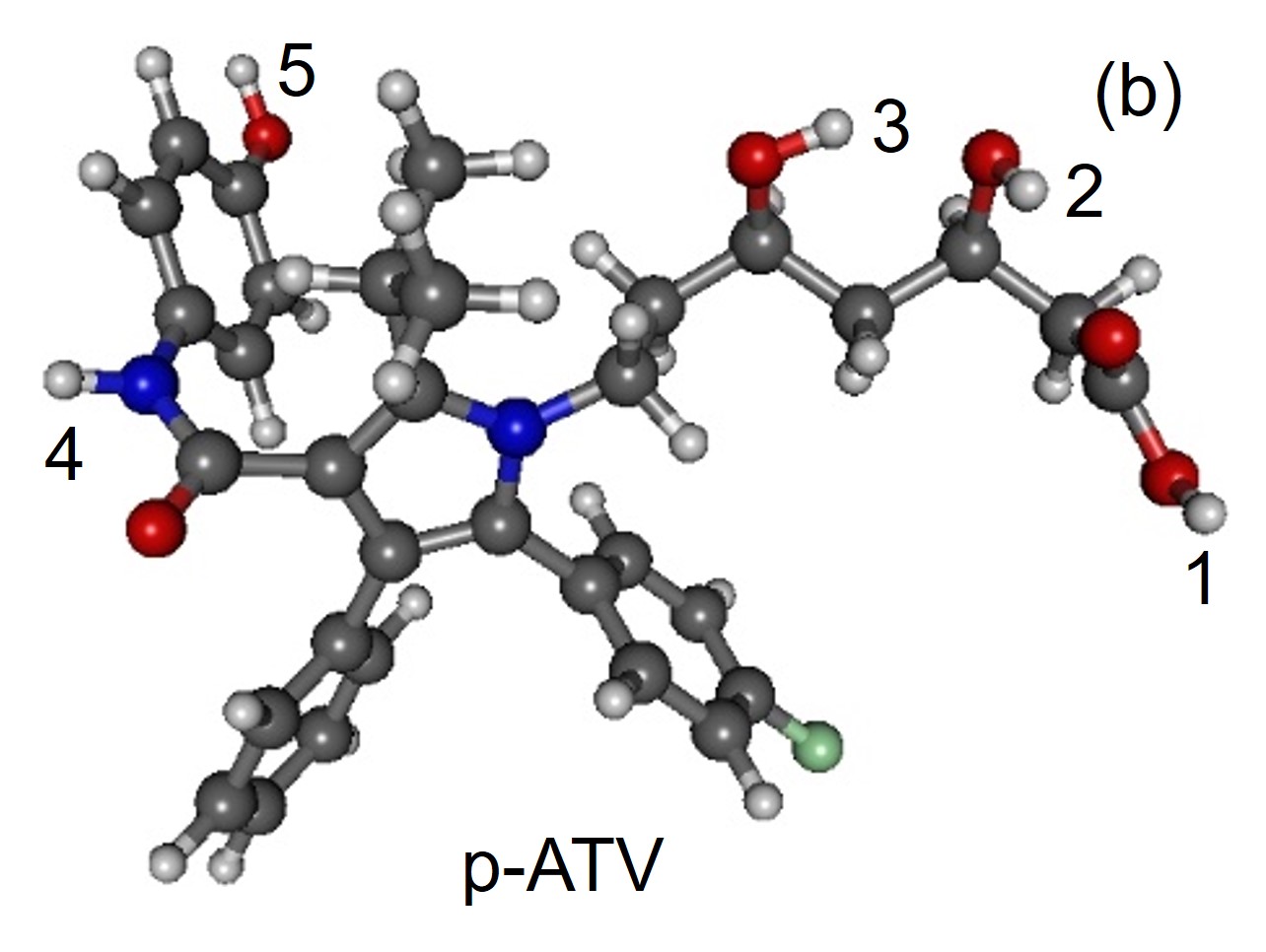}
}

\caption{Optimized geometries of atorvastatin ortho- and para-hydroxy metabolites: (\textbf{a}) o-ATV and (\textbf{b}) p-ATV.
 Figure generated using GABEDIT \cite{gabedit}.\label{fig:geom-oatv-patv}}
\end{figure}%mdpi: Figure 2 is not cited in this article, plaese confirm and modify.

\vspace{-6pt}
\begin{figure}[H]
{
\includegraphics[width=5.5cm]{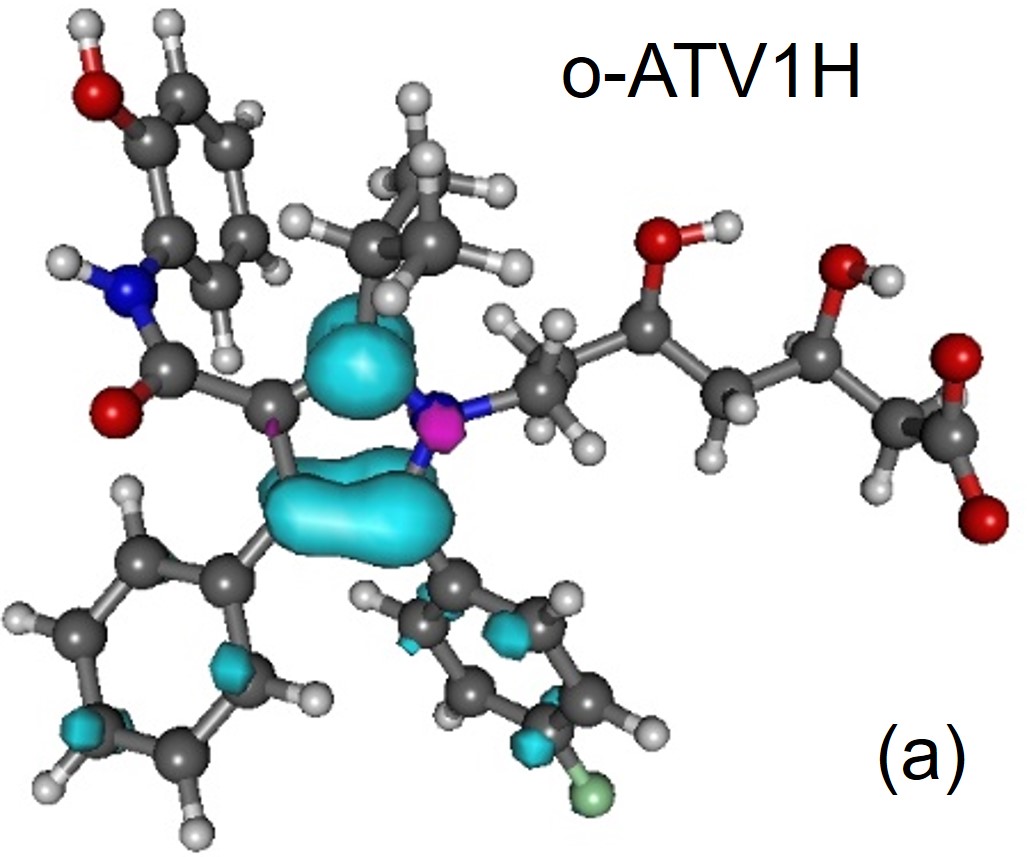}
\includegraphics[width=6.3cm]{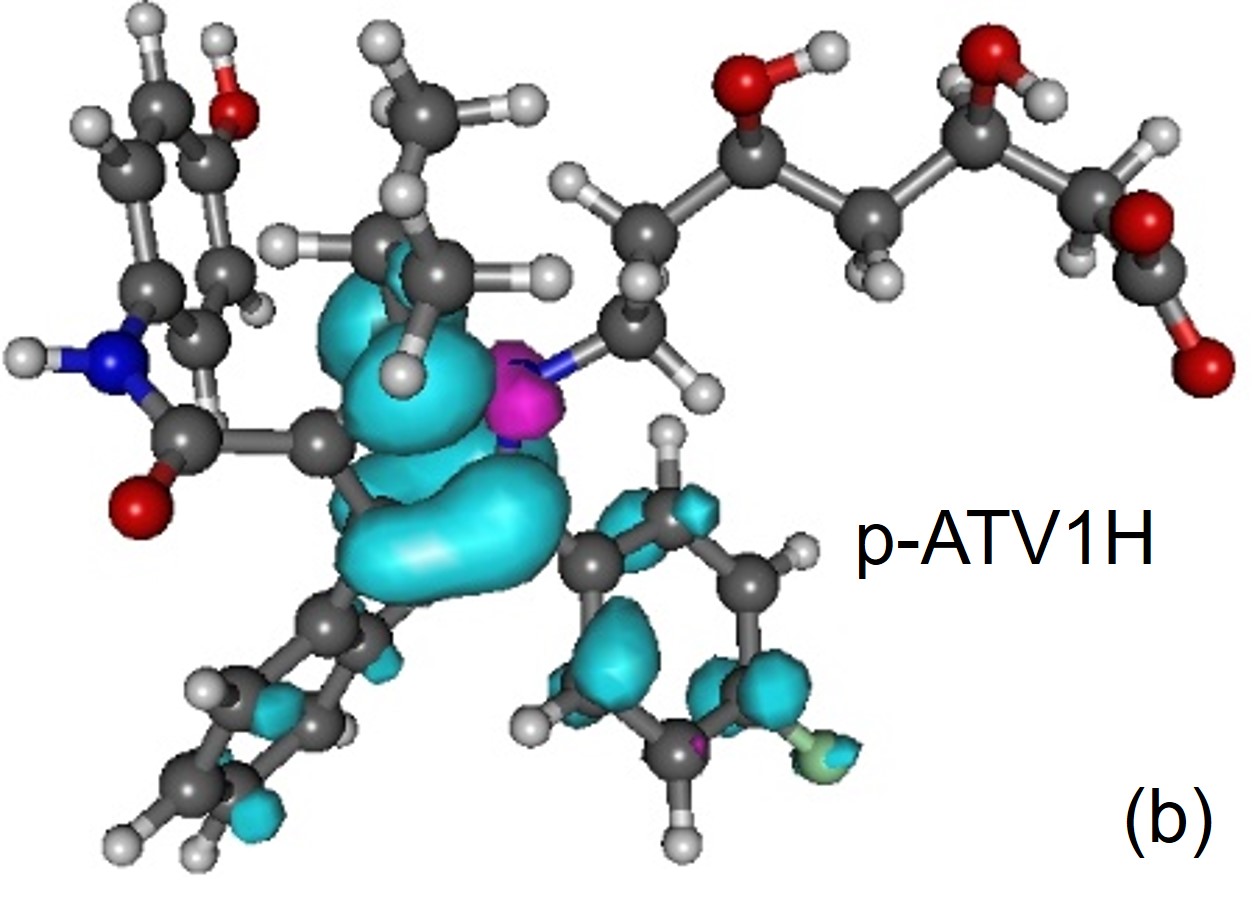}
}

\caption{Spin densities of radicals generated by H-atom abstraction at position 1-OH:
  (\textbf{a}) o-ATV1H and (\textbf{b}) p-ATV1H.
  Figure generated using GABEDIT \cite{gabedit}.\label{fig:1-oh-oatv-patv}}
\end{figure}
\vspace{-6pt}
\begin{figure}[H]

{
\includegraphics[width=6.2cm]{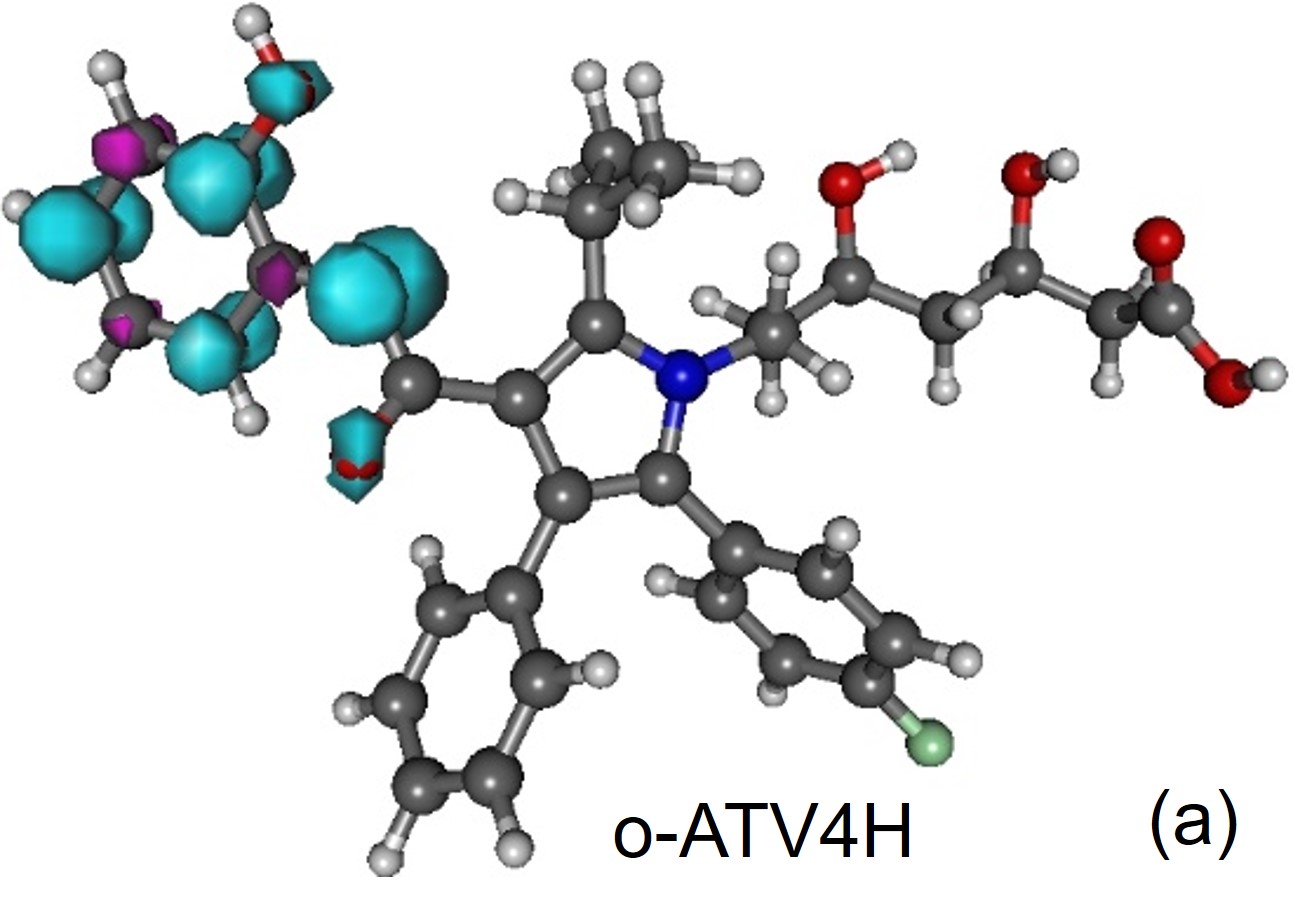}
\includegraphics[width=6.7cm]{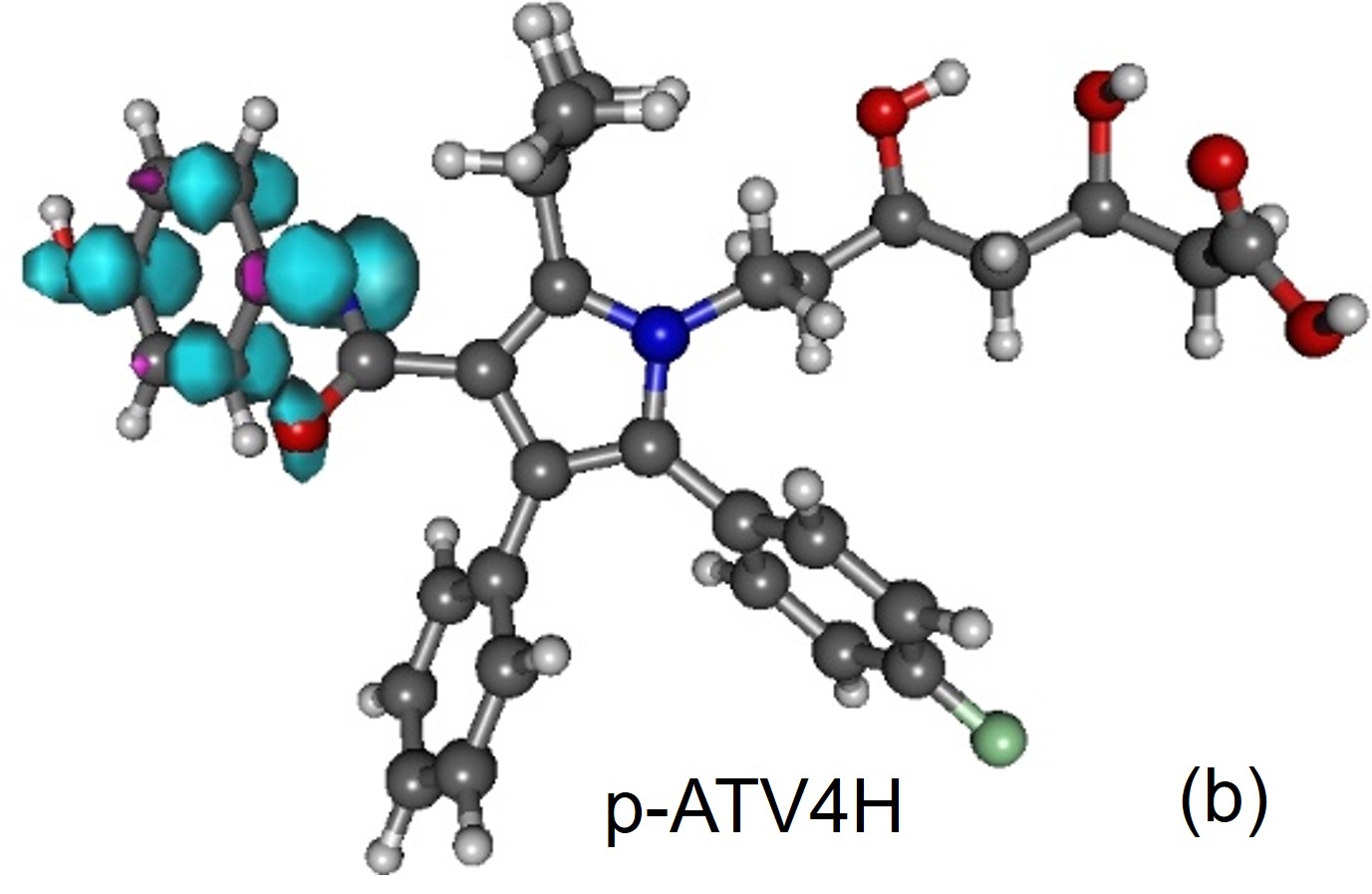}
}

\caption{
Spin densities of radicals generated from atorvastatin ortho- and para-hydroxy metabolites by H-atom abstraction at position 4-NH:
(\textbf{a}) o-ATV4H and (\textbf{b}) p-ATV4H.
Figure generated using GABEDIT \cite{gabedit}.\label{fig:4-nh-oatv-patv}}
\end{figure}

\vspace{-6pt}
\begin{figure}[H]

{
\includegraphics[width=6cm]{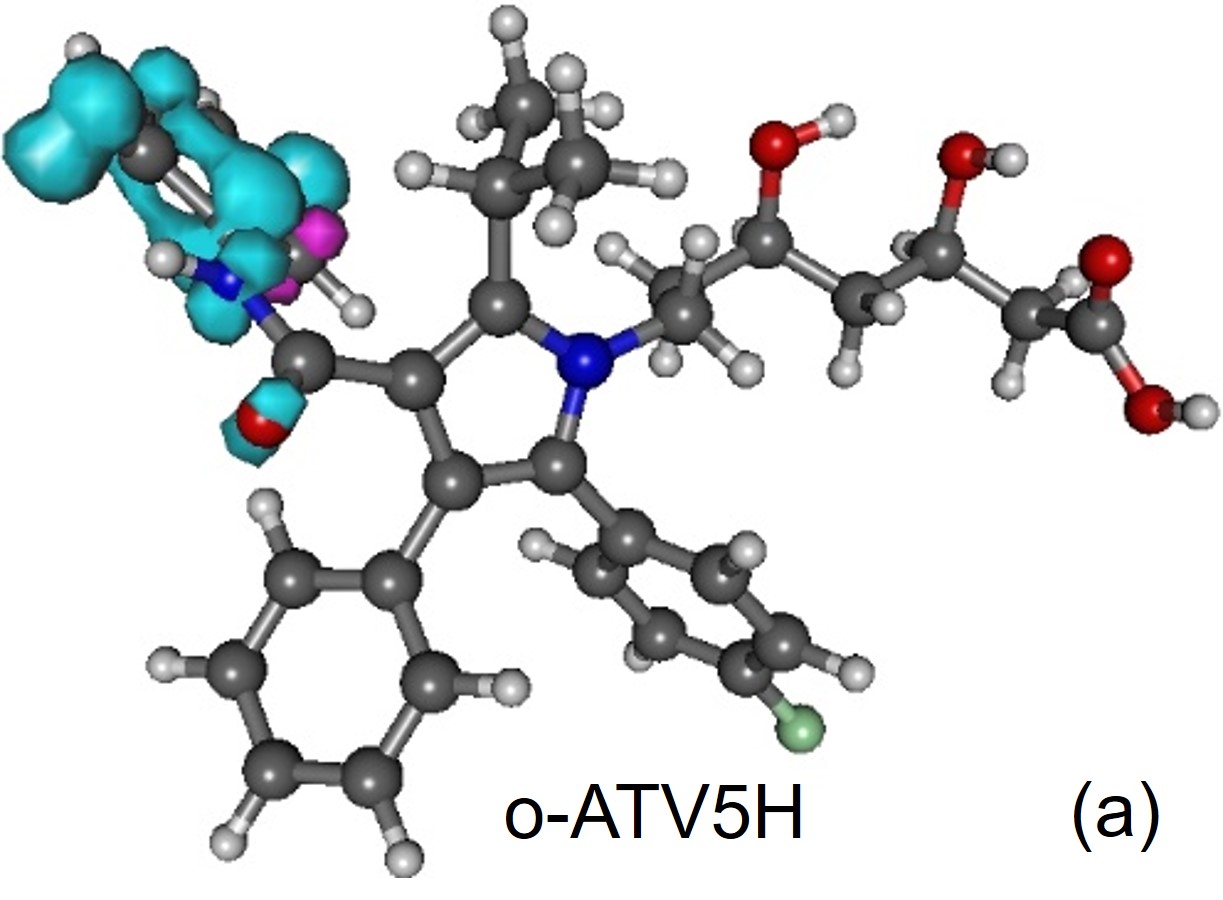}
\includegraphics[width=6cm]{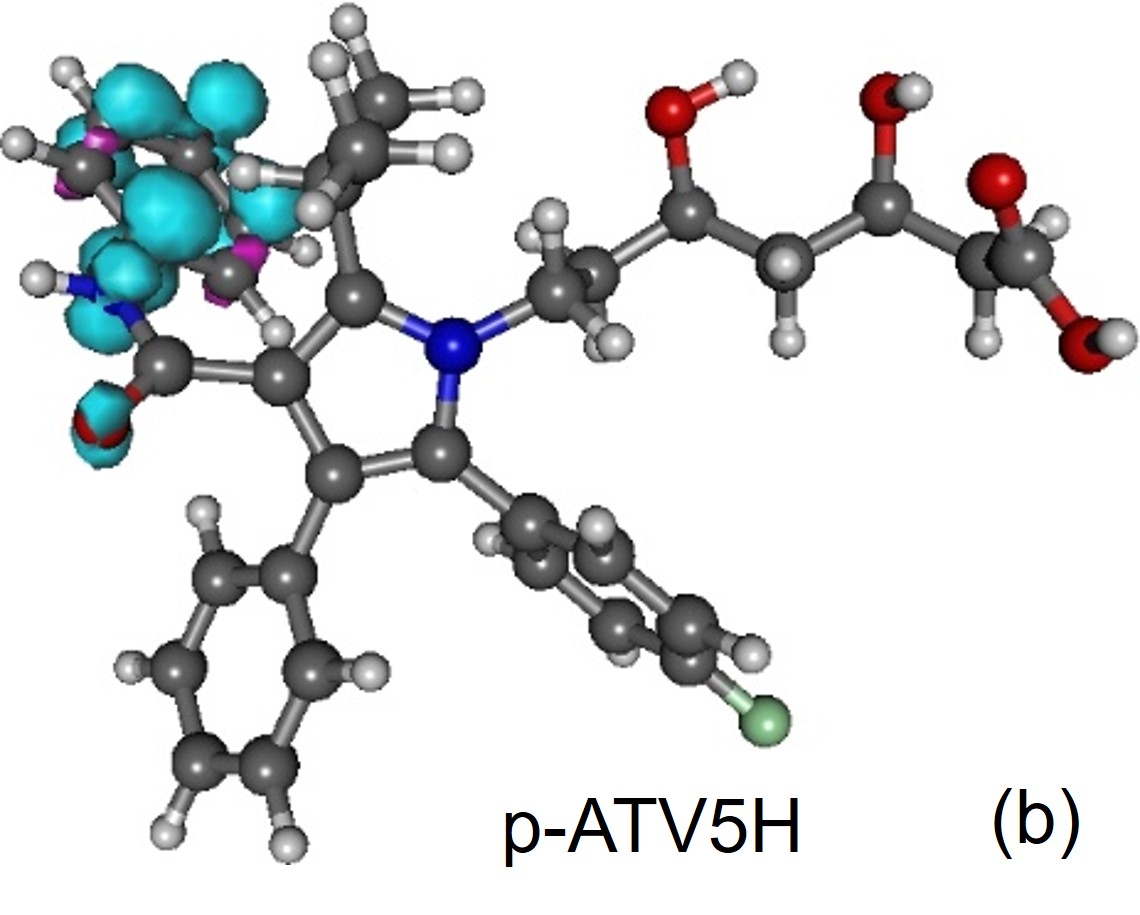}
}

\caption{
Spin densities of radicals generated from atorvastatin ortho- and para-hydroxy metabolites by H-atom abstraction at position 5-OH:
(\textbf{a}) o-ATV5H and (\textbf{b}) p-ATV5H.
Figure generated using GABEDIT \cite{gabedit}.\label{fig:5-oh-oatv-patv}}
\end{figure}

\section{Results and Discussion}
\label{sec:results}
\subsection{Molecular Geometries}
\label{sec:geometry}
Along with the neutral, cation and anion ATV 
--- molecular formula \ce{C33H35FN2O5}, IUPAC name
(3R,5R)-7-[2-(4-fluorophenyl)-3-phenyl-4-(phe\-nyl\-car\-ba\-moyl)-5-pro\-pan-2-yl\-pyr\-rol-1-yl]-3,5-di\-hy\-droxy\-hep\-tanoic acid,
CAS number 134523-00-5--- and its metabolites ortho-hy\-dro\-xy ator\-vas\-ta\-tin
(o-ATV and para--hy\-dro\-xy\-ator\-vas\-ta\-tin (p-ATV),
we also investigated the radicals (e.g., ATV$n$H$^\bullet$) generated by H-atom abstraction from their \ce{O-H} and \ce{N-H} groups
as well as the anions \ce{ATV$n$H-} of the latter. Here, $n(=1,2,3,\ldots)$ labels the various positions of the H-atoms, as depicted in
{Figures} \ref{fig:atv}, \ref{fig:geom-oatv-patv}, \ref{fig:1-oh-oatv-patv}, \ref{fig:4-nh-oatv-patv} and \ref{fig:5-oh-oatv-patv}.

All quantities to be discussed below were calculated at the total electronic energy minima of the species listed above
obtained via B3LYP/6-31+G(d,p)/IEFPCM optimization (cf.~\secname\ref{sec:methods}), which
(with the grain of salt mentioned in \secname\ref{sec:nbo}) % the caption of {Figure} \ref{fig:ir}) 
posed no special problems.
Neither H-atom abstraction ({Figures} \ref{fig:atv}b, c, and d)
nor ortho- and para-\ce{O-H} addition ({Figures} \ref{fig:geom-oatv-patv} a and b)
spectacularly modifies the molecular conformation (({Figures} \ref{fig:atv}a).
$Z$-matrices for optimized geometries of representative species are presented in Tables~\ref{table:zmat-atv}--\ref{table:zmat-atv1h-elements} and
{Figures} \ref{fig:atv}, \ref{fig:1-oh-oatv-patv}, \ref{fig:4-nh-oatv-patv} and \ref{fig:5-oh-oatv-patv}.
Rather than Cartesian coordinates, we prefer to show $Z$-matrices because they facilitate comparison between
various species and methods.

\subsection{Chemical Reactivity Indices}
\label{sec:eta}
The global chemical reactivity indices investigated in this work are listed below along with their expressions
in terms of the ionization potential IP and electroaffinity EA \cite{Parr:89,Gazquez:07,Rajan:18,Baldea:2019f,Baldea:2019g}:
\begin{equation}
\begin{array}{ll}
    \mbox{\small chemical hardness} &  \eta \equiv E_g/2,  \\
    \mbox{\small chemical softness} & \sigma \equiv 1/E_g , \\
    \mbox{\small electronegativity} & \chi \equiv (\mbox{IP} + \mbox{EA})/2 , \\
    \mbox{\small electrophilicity index} & \omega \equiv \chi^2/(2 \eta) , \\
    \mbox{\small electroaccepting power} & \omega^{+} \equiv  (\mbox{IP} + 3\, \mbox{EA})^2/(16 E_g) , \\
    \mbox{\small electrodonating power} & \omega^{-} \equiv (3 \,\mbox{IP} + \mbox{EA})^2/(16 E_g) .\\
\end{array}
\label{eq-eta}
\end{equation}
Here, $E_g \equiv  \mbox{IP} - \mbox{EA}$ is the fundamental (or transport) ``HOMO-LUMO''
gap \cite{Parr:89,Burke:12,Baldea:2014c}.
Noteworthily, the values of IP and EA presented in this paper were calculated as enthalpy differences (cf.~\gls~(\ref{eq-setpt-ip}) and (\ref{eq-ea})).
Estimating IP and EA using the eigenvalues of the Kohn-Sham (KS) orbitals with reversed sign (Koopmans theorem),
\begin{equation}
  IP \to I = - E_{HOMO}^{KS}; \ EA \to A = - E_{HOMO}^{KS},
  \label{eq-i-a}
\end{equation}
is unfortunately a very popular approximation, but it is totally inadequate especially in the presence of a solvent.
For clarification, a comment is in order at this point.

  Although both the approach using \gl~(\ref{eq-i-a}) and the approach using \gls~(\ref{eq-setpt-ip}) and (\ref{eq-ea})
  are based on the DFT, there is an important difference between them.  

  \Gls~(\ref{eq-setpt-ip}) and (\ref{eq-ea}) rely on total energies computed via DFT. In these computations,
  the Kohn-Sham (KS) orbitals merely enter as eigenfunctions and eigenvalues of a mathematical (minimization) problem.
  They are auxiliary mathematical objects useful to compute a quantity with a clear physical meaning
  (namely, the total electronic energy).

  However, it should be well known to any well rounded theoretician
  that the KS ``orbitals''
  do not have any physical meaning; they are \emph{not} real molecular orbitals \cite{Parr:89,Kohn:96,Baldea:2014c}.
  What makes \gl~(\ref{eq-i-a}) problematic is just the fact that it treats the KS-HOMO and KS-LUMO \emph{as if} they were
  the true HOMO and LUMO of a real molecule. 
  
  To remedy the difficulty related to the KS ``energies''
  (in reality, eigenvalues of a mathematical single-particle problem) in semiconductor physics, which translates into KS-band gaps
  typically amounting to about 50\% of the real
  band gap, a so-called ``scissor'' operator procedure is applied \cite{Godby:88,Baldereschi:95}, which consists of empirically shifting the KS
  eigenvalues.
  To eliminate this severe difficulty in the case of molecules immersed in solvents, we also proposed a scissor technique \cite{Baldea:2013c}.
  The important difference is that the scissor corrections proposed in ref.~\cite{Baldea:2013c} are obtained from quantum chemical calculations
  rather than empirically as done in semiconductor band structure calculations.

Switching back, one may expect that the global chemical reactivity indices can give a flavor of the overall stability of a molecule
and are useful in predicting how a certain chemical environment evolves in time \cite{Baldea:2020c,Baldea:2020e}.
In certain situations they turned out to be useful for comparing properties of different molecular species \cite{Parr:99,Gazquez:07,Gazquez:08}.

\begin{figure}[H]
{
\hspace{-6pt}\includegraphics[width=6.5cm]{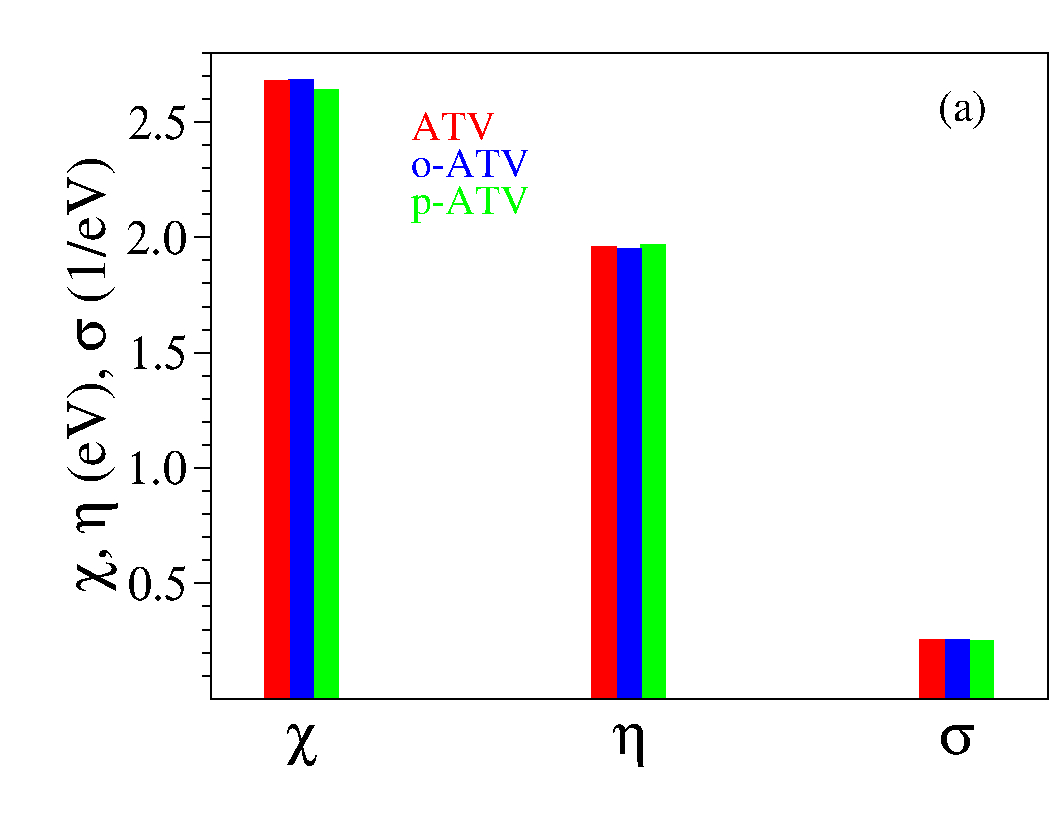}
\includegraphics[width=6.5cm]{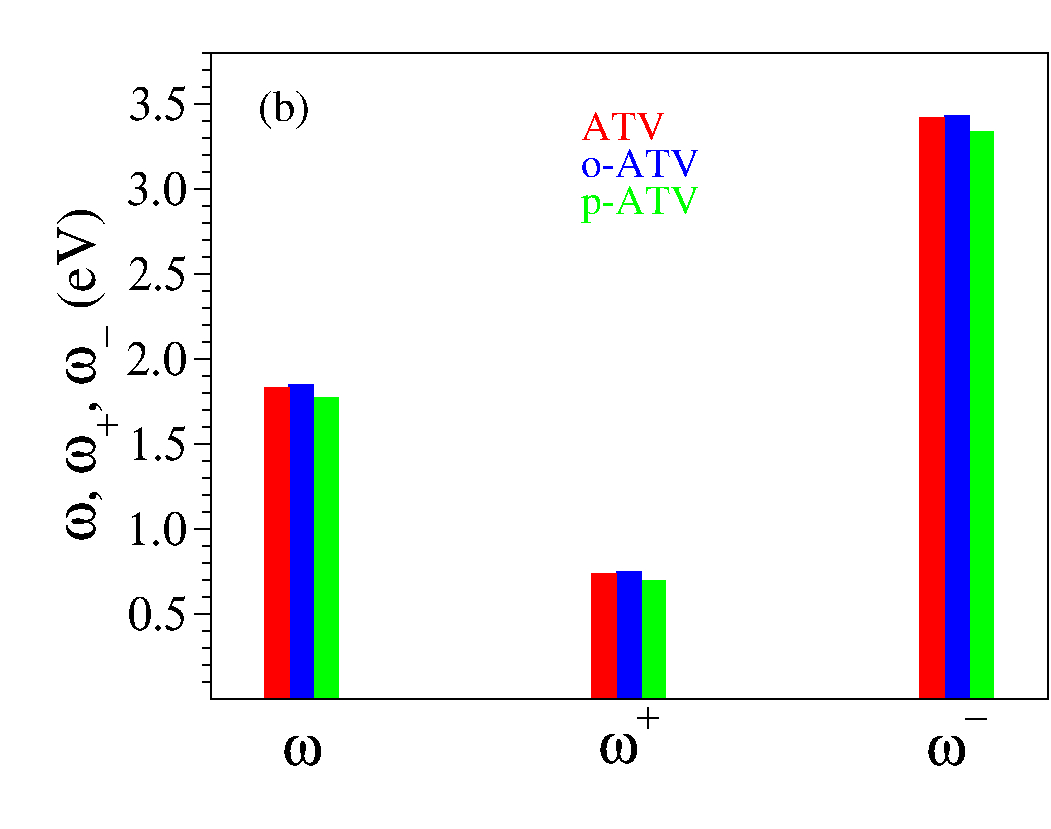}
}

\caption{Global chemical reactivity indices defined by \gl~(\ref{eq-eta}) for
  atorvastatin and its ortho- and para-hydroxy metabolites: (a) electronegativity $\chi$, chemical hardness $\eta$, and chemical softness
  $\sigma$; (b) electrophilicity index $\omega$, electroaccepting power $\omega^{+}$, and electrodonating power $\omega^{-}$. 
  \label{fig:eta-etc}}
\end{figure}

The presently calculated global chemical reactivity indices of ATV and its metabolites
are collected in Tables~\ref{table:eta-methanol} and \ref{table:eta-methods}, and depicted in Figure \ref{fig:eta-etc}.
Having a chemical hardness $\eta$ of about 2\,eV,
ATV, o-ATV, and p-ATV exhibit a good chemical stability. This value lies between the values of the natural antioxidants phenol and trolox,
for which our calculations at the same B3LYP/6-31+G(d,p)/IEFPCM level
yielded $\eta = 2.56$\,eV and $\eta = 1.88$\,eV, respectively.
For all three species, the electrophilic indices \cite{Parr:99,Gazquez:07,Gazquez:08}
are $\omega \approx 1.8$\,eV, a value exceeding the value of 1.50\,eV, which is considered the threshold for strong electrophiles \cite{Domingo:02}.
For comparison, let us again mention the values $\omega = 1.61$\,eV and $\omega = 1.85$\,eV computed by us
for phenol and trolox, respectively.

Inspection of Table~\ref{table:eta-methanol} reveals that, similarly to the quantities $\eta$ and $\omega$
considered above, all global chemical reactivity of ATV, o-ATV, and p-ATV are comparable to those of
well known natural antioxidants. Could we then expect that 
ATV flavors (or other molecules) have indeed good antioxidant potency merely based on
global chemical reactivity indices comparable to those of good antioxidants?

\begin{table}[H]
\small
  \setlength{\tabcolsep}{3.56mm}
  
    \caption{Global chemical reactivity indices (eV) computed via B3LYP/6-31+G(d,p)/IEFPCM
      for atorvastatin and its ortho- and para-hydroxy metabolites and two natural oxidants in methanol.}                                    
  \label{table:eta-methanol}  
  %%%%%%%%%%%%%%%%%%%%%%%%%%%%
    \begin{tabular*}{\textwidth}{@{\extracolsep{\fill}}lccccccccc}
      \toprule
\textbf{Molecule}                  & \textbf{IP}         &  \textbf{EA}         &     \boldmath{$E_g$ }      &      \boldmath{ $\eta$} & \boldmath{$\mu$}  & \boldmath{$\sigma$} & \boldmath{$\omega $} & \boldmath{$\omega^{+}$} &\boldmath{ $\omega^{-}$}  \\
      \midrule
ATV                       &        4.64  &        0.72   &   3.92         &       1.96   &       $-$2.68  &         0.26     &         1.83      &            0.74    &          3.42       \\
o-ATV                  &        4.64  &        0.73   &   3.90         &       1.95   &       $-$2.68  &         0.26     &         1.85      &            0.75    &          3.43       \\ 
p-ATV                  &        4.60  &        0.67   &   3.93         &       1.97   &       $-$2.64  &         0.25     &         1.77      &            0.70    &          3.34       \\[0.5ex]
    \midrule                                                                                                        
Phenol                    &        5.43  &        0.31   &   5.12         &      2.56   &       $-$2.87  &          0.20    &          1.61     &           0.49     &          3.36       \\
Trolox                    &        4.51  &        0.76   &   3.76         &      1.88   &       $-$2.63  &          0.27    &          1.85     &           0.77     &          3.40       \\
    \bottomrule                     
    \end{tabular*}                                                                                         
    %%%%%%%%%%%%%%%%%%%%%%%%%%%%%%%%%%%%%%%%%%%%%%%%%%%%%%%%%%%%%%%%%%%%%%%%%%%%%%%%%%%%%%                
  %%%%%%%%%%%%%%%%%%%%%%%%%%%%%%%%%%%%%%%%%%%%%%%%%%%%%%%%%%%%%%%%%%%%%%%%%%%%%%%%%%%%%%                   

\end{table}                                                                                               
\vspace{-9pt}

\begin{table}[H]
\small
  \setlength{\tabcolsep}{3.56mm}
    \caption{Global chemical reactivity indices (eV) for ATV in methanol computed using 6-31+G(d,p) basis sets 
and the exchange-correlation functionals (B3LYP, PBE0, M062x) and solvent models (IEFPCM, SMD) specified above.} 
  \label{table:eta-methods}                                                                                        
    
  %%%%%%%%%%%%%%%%%%%%%%%%%%%%
  \begin{adjustwidth}{-\extralength}{0cm}%\centering
  \begin{tabular*}{\fulllength}{@{\extracolsep{\fill}}clccccccccc}
      \toprule
\textbf{Molecule}  & \textbf{Method}                  & \textbf{IP}         &   \textbf{EA}         &    \boldmath{$E_g$ }      &     \boldmath{  $\eta$} &\boldmath{ $\mu$}  &\boldmath{ $\sigma$} & \boldmath{$\omega $ }&\boldmath{ $\omega^{+}$} & \boldmath{$\omega^{-}$}  \\
      \midrule
ATV   &    UB3LYP/IEFPCM       &        4.64  &        0.72   &   3.92         &       1.96   &       $-$2.68  &         0.26     &         1.83      &            0.74    &          3.42       \\
      &    B3LYP/SMD      &    4.39 &    0.63  & 3.76      &   1.88  &   $-$2.51 &     0.27    &     1.67     &        0.65   &      3.16      \\
      &    ROB3LYP/IEFPCM      &        4.69  &        0.70   &   3.98         &       1.99   &       $-$2.69  &         0.25     &         1.82      &            0.72    &          3.42       \\
      &    UPBE0/IEFPCM        &        4.67  &        0.71   &   3.96         &       1.98   &       $-$2.69  &         0.25     &         1.82      &            0.73    &          3.41       \\
      &    UM062x/IEFPCM  &   4.95  &   0.73   &4.22       &   2.11  &   $-$2.84 &     0.24    &     1.91     &        0.75   &      3.59      \\
      \midrule                                                                                                                                                                      
o-ATV &    UB3LYP/IEFPCM       &        4.64  &        0.73   &   3.90         &       1.95   &      $-$2.68  &         0.26     &         1.85      &            0.75    &          3.43       \\ 
      &    UB3LYP/SMD     &    4.38 &    0.63  &3.75       &   1.87  &   $-$2.51 &     0.27    &     1.68     &        0.66   &       3.16     \\    
      \midrule                                                                                                                                                                    
p-ATV &    UB3LYP/IEFPCM    &        4.60  &        0.67   &   3.93         &       1.97   &       $-$2.64  &         0.25     &         1.77      &            0.70    &          3.34       \\
      &    UB3LYP/SMD     &    4.37 &    0.61  &3.77       &   1.88  &   $-$2.49 &     0.27    &     1.65     &        0.64   &       3.13     \\       
 \bottomrule                        
    \end{tabular*}        
\end{adjustwidth}                                                                                 
    %%%%%%%%%%%%%%%%%%%%%%%%%%%%%%%%%%%%%%%%%%%%%%%%%%%%%%%%%%%%%%%%%%%%%%%%%%%%%%%%%%%%%%                
  %%%%%%%%%%%%%%%%%%%%%%%%%%%%%%%%%%%%%%%%%%%%%%%%%%%%%%%%%%%%%%%%%%%%%%%%%%%%%%%%%%%%%%                   
                                                                                            
\end{table}                                                                                               
\unskip 
%mdpi: there is no explanation of the subfigure (a) and (b), plaese confirm and add it.

The analysis in the next section will unravel that, in fact, the global chemical indices have little relevance
for assessing the antioxidant activity of a certain molecule. For the time being, let us remark that
the values of Table~\ref{table:eta-methanol} would rather suggest that ATV and o-ATV have similar 
antioxidant properties, and that ATV (possibly) performs (slightly) better than p-ATV.

\subsection{Antioxidant Mechanisms and Pertaining Enthalpies of Reaction}
\label{sec:bde}
As is widely discussed in the literature, an H-atom can be transferred to a free radical in one or two step processes.
The three antioxidative mechanisms (HAT, SET-PT, and SPLET)
and the corresponding reaction enthalpies (BDE, IP and PDE, PA and ETE, respectively) can be expressed as follows:

Direct hydrogen atom transfer (HAT) \cite{Burton:85,deHeer:00,Mayer:04}
\begin{equation}
  \label{eq-hat}
 \begin{array}{cc}
   \ce{{AX}H + R^{\bullet}} \to \ce{{AX}^{\bullet}} + \ce{RH} &
       \mbox{\small BDE} = H\left(\ce{{AX}^{\bullet}}\right) + H\left(\ce{H^{\bullet}}\right) - H\left(\ce{{AX}H}\right) .
 \end{array}
\end{equation}

 Stepwise electron transfer proton transfer (SET-PT) \cite{Jovanovic:94,Jovanovic:96}
 \begin{subequations}
  \label{eq-setpt}
   \begin{equation}
   \label{eq-setpt-ip}
   \begin{array}{cc}
     \ce{{AX}H} \to \ce{{AX}H^{\bullet +}} + \ce{e-} &
     \mbox{\small IP} = H\left(\ce{{AX}H^{\bullet +}}\right) + H\left(\ce{e-}\right) - H\left(\ce{{AX}H}\right) .
      \end{array}
 \end{equation}
 \begin{equation}
     \label{eq-setpt-pde}
   \begin{array}{cc}
     \ce{{AX}H^{\bullet +}} \to  \ce{{AX}^{\bullet}} + \ce{H+} &
       \mbox{\small PDE} = H\left( \ce{{AX}^{\bullet}}\right) + H\left(\ce{H+}\right) -  H\left(\ce{{AX}H^{\bullet +}}\right) .
   \end{array}
 \end{equation}
 \end{subequations}

 Sequential proton loss electron transfer (SPLET) \cite{Litwinienko:03,Litwinienko:04}
 \begin{subequations}
   \label{eq-splet}
   \begin{equation}
     \label{eq-splet-pa}
   \begin{array}{cc}
     \ce{{AX}H} \to \ce{{AX}-} + \ce{H+} &
      \mbox{\small PA} = H\left(\ce{{AX}-}\right) + H\left(\ce{H+}\right) - H\left(\ce{{AX}H}\right)
      \end{array}
 \end{equation}
 \begin{equation}
       \label{eq-splet-ete}
   \begin{array}{cc}
     \ce{{AX}-} \to  \ce{{AX}^{\bullet}} + \ce{e-} &
       \mbox{\small ETE} = H\left( \ce{{AX}^{\bullet}}\right) + H\left(\ce{e-}\right) -  H\left(\ce{{AX}-}\right) .
   \end{array}
 \end{equation}
 \end{subequations}

In our specific case, {X} stands for an O or an N atom.

Related to the above (albeit not directly entering the aforementioned antioxidation mechanisms), the electron attachment
process is quantified by the electroaffinity defined as
\begin{equation}
  \label{eq-ea}
  \mbox{\small EA} = H\left(X\right) + H\left(\ce{e-}\right) - H\left(\ce{X-}\right) .
\end{equation}
BDE, IP, PDE, PA, and ETE are enthalpies of reaction which can be obtained as adiabatic properties 
from standard $\Delta$-DFT prescriptions
\cite{Gunnarson:89,Baldea:2012i,Baldea:2014c}. To this aim, along with the enthalpies of the various ATV-based species entering the 
above reactions, the enthalpies of the H-atom, proton and electron in methanol are also needed \cite{baldea_2022_chemrxiv}. 
They are presented in Table~\ref{table:values}.
\begin{table}[H]
\small
  \setlength{\tabcolsep}{6.56mm}
  
  \caption{Gas phase enthalpies $H_0$ and solvation enthalpies $\Delta H_{sol}$
      in hartree utilized in the present calculations. For the for the gas phase enthalpy
      of the H-atom we used the value for the B3LYP/6-31+G(d,p) electronic energy ($-0.500273$\,hartree)
      and the value of 1.4816\,kcal/mol for thermal correction to enthalpy common for all compound model chemistries from GAUSSIAN 16.}
  \label{table:values}
    %%%%%%%%%%%%%%%%%%%%%%%%%%%%
    \begin{tabular*}{\textwidth}{@{\extracolsep{\fill}}lcccccc}
      \toprule
     \textbf{ Species } &      \boldmath{ $H_0$ }     &  \boldmath{$\Delta H_{sol}^{methanol}$} \\
      \midrule
      Electron & $+$0.001194\mbox{ $^a$}\hspace*{-1.5ex}      & $-$0.030204\mbox{ $^c$}\hspace*{-1.5ex}  \\
      Proton   & $+$0.002339\mbox{ $^b$}\hspace*{-1.5ex}      & $-$0.405508\mbox{ $^c$}\hspace*{-1.5ex}  \\
      H-atom   & $-$0.497912 % -0.500273                                  
                                                             & $+$0.001904\mbox{ $^d$}\hspace*{-1.5ex}  \\
      \bottomrule
      %%%%%%%%%%%%%%%%%%%%%%%%%%%%%%%%%%%%%%%%%%%%%%%%%%%%%%%%%%%%%%%%%%%%%%%%%%%%%%%%%%%%%%
    \end{tabular*}
    %%%%%%%%%%%%%%%%%%%%%%%%%%%%%%%%%%%%%%%%%%%%%%%%%%%%%%%%%%%%%%%%%%%%%%%%%%%%%%%%%%%%%%
    \noindent{\footnotesize{\textsuperscript{a} From Ref.~\cite{Fifen:13}. \textsuperscript{b}   From Ref.~\cite{Fifen:14}.  \textsuperscript{c} Form Ref.~\cite{Markovic:16}.  \textsuperscript{d} Form Ref.~\cite{Rimarcik:10}.  }}%mdpi: there is no superscript a--d in the table, please confirm and modify.
% 
    %%%%%%%%%%%%%%%%%%%%%%%%%%%%%%%%%%%%%%%%%%%%%%%%%%%%%%%%%%%%%%%%%%%%%%%%%%%%%%%%%%%%%%
\end{table}

\begin{figure}[H]

\centerline{\includegraphics[width=6.5cm]{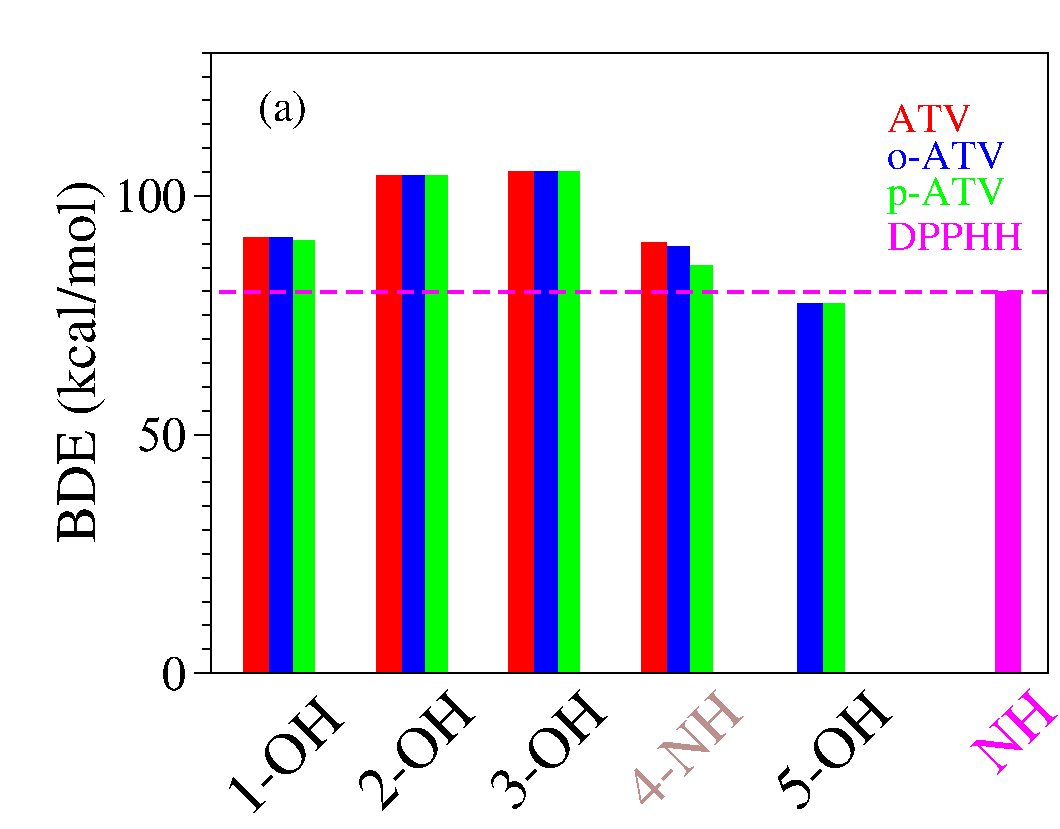}}

\includegraphics[width=6.5cm]{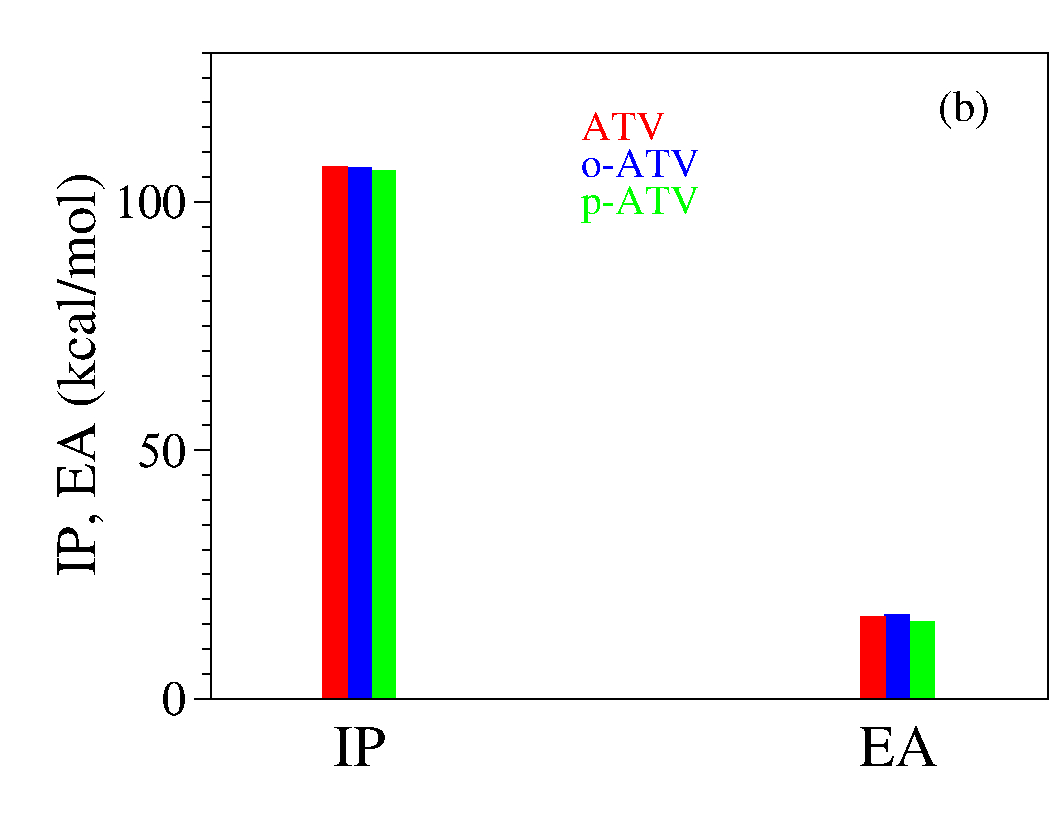}
\includegraphics[width=6.5cm]{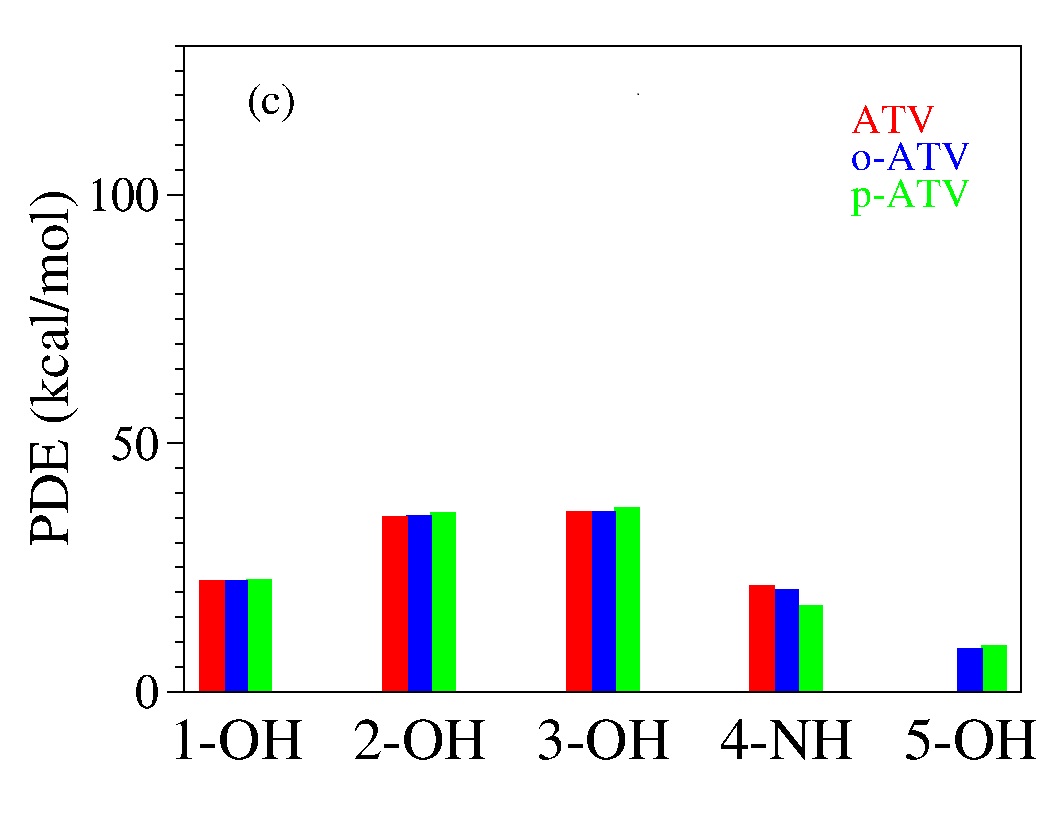}

\includegraphics[width=6.5cm]{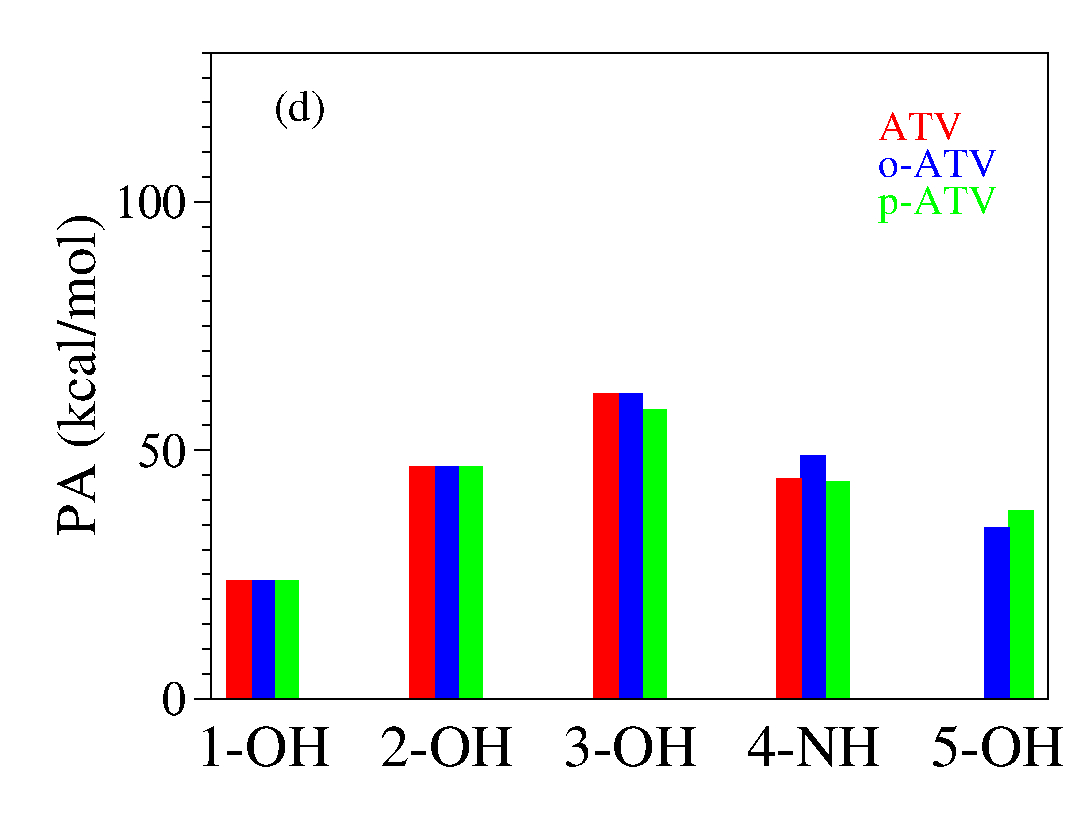}\includegraphics[width=6.5cm]{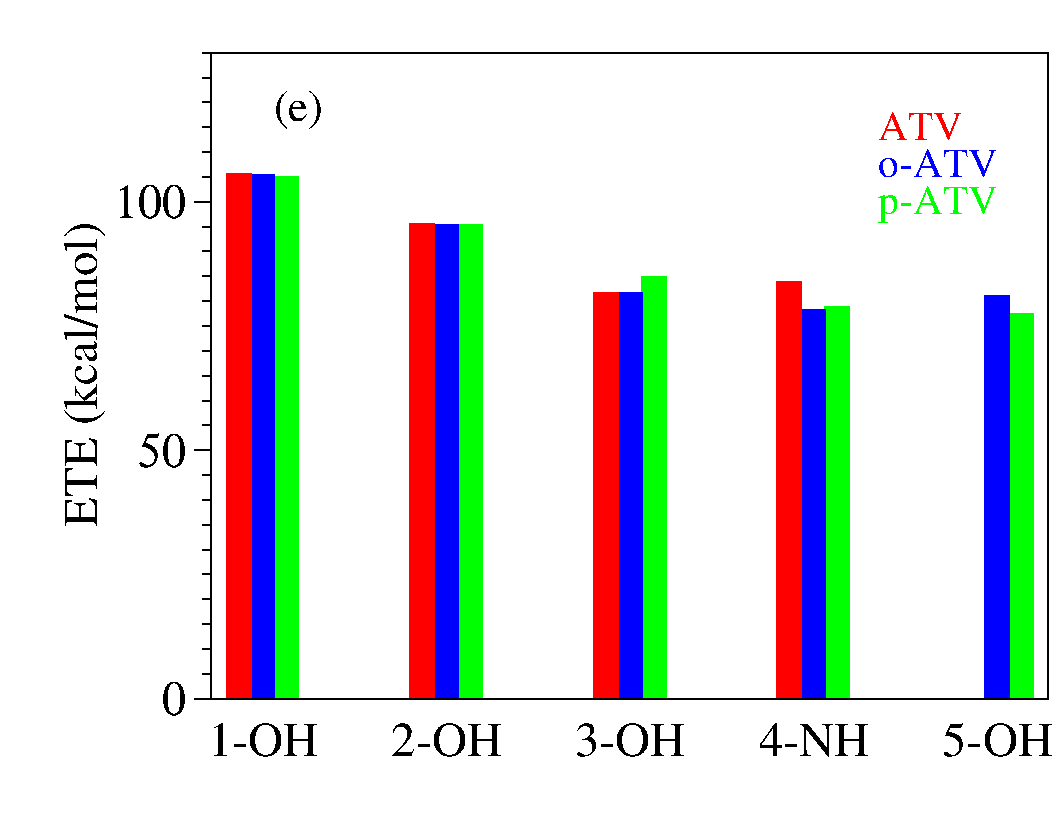}

\caption{
Enthalpies of reaction quantifying the antioxidant activity of atorvastatin (ATV) and its ortho- (o-ATV) and para- (p-ATV) hydroxy metabolites:
(\textbf{a}) bond dissociation; (\textbf{b}) ionization and electron attachment; (\textbf{c}) proton detachment; (\textbf{d}) proton affinity; (\textbf{e}) electron transfer. The additional information 
for the DPPH$^\bullet$ radical in panel (\textbf{a}) depicts 
why o-ATV and p-ATV can scavenge this radical while the parent ATV cannot.\label{fig:bde-etc}}
\end{figure}

The presently computed thermodynamic parameters are collected in Tables~\ref{table:bde} and \ref{table:bde-pbe0},
and depicted in {Figure} \ref{fig:bde-etc}. 
\begin{table}[H]
 \small
  \setlength{\tabcolsep}{5.56mm}
   \caption{The enthalpies of reaction (in kcal/mol) needed to quantify the antioxidant activity of atorvastatin
      (ATV) and its ortho- and para-hydroxy metabolites (o-ATV, p-ATV).}
  \label{table:bde}

    \begin{tabular*}{\textwidth}{@{\extracolsep{\fill}}lcccccc}
      \toprule
  \textbf{ Molecule}              & \textbf{Position}   &   \textbf{BDE}            &         \textbf{IP}   &        \textbf{PDE}     &        \textbf{PA}    &       \textbf{ ETE}   \\  
      \midrule                                        
{ATV}                    &  1-OH      &           91.4   &       107.0  &        22.4    &        23.8  &       105.6  \\
                         &  2-OH      &          104.2   &              &        35.3    &        46.7  &        95.6  \\
                         &  3-OH      &          105.2   &              &        36.3    &        61.5  &        81.8  \\
                         &  4-NH      &           90.2   &              &        21.3    &        44.4  &        83.9  \\
\midrule                                                                              
{o-ATV   }               &  1-OH      &           91.2   &       106.9  &        22.4    &        23.8  &       105.5  \\
                         &  2-OH      &          104.2   &              &        35.4    &        46.8  &        95.5  \\
                         &  3-OH      &          105.1   &              &        36.3    &        61.5  &        81.7  \\
                         &  4-NH      &           89.3   &              &        20.5    &        49.0  &        78.4  \\
                         &  5-OH      &           77.5   &              &         8.7    &        34.4  &        81.2  \\
     \midrule                                                            
{p-ATV}                  &  1-OH      &           90.7   &       106.2  &        22.6    &        23.8  &       105.0  \\
                         &  2-OH      &          104.2   &              &        36.0    &        46.8  &        95.4  \\
                         &  3-OH      &          105.1   &              &        37.0    &        58.2  &        85.0  \\
                         &  4-NH      &           85.5   &              &        17.4    &        43.8  &        79.0  \\
                         &  5-OH      &           77.4   &              &         9.2    &        37.9  &        77.5  \\
     \bottomrule                                                          

    \end{tabular*}
  %%%%%%%%%%%%%%%%%%%%%%%%%%%%%%%%%%%%%%%%%%%%%%%%%%%%%%%%%%%%%%%%%%%%%%%%%%%%%%%%%%%%%%
   
\end{table}

\begin{table}[H] % [h!]
  \small
  \setlength{\tabcolsep}{4mm}
   \caption{Enthalpies of reaction (in kcal/mol) computed for atorvastatin (ATV) using methods indicated above and 6-31+G(d,p) basis sets.
      There is no difference between
    unrestricted (UB3LYP) and restricted open shell (ROB3LYP) methods in calculating the PA values, and for this reason the pertaining value was written in parenthesis.}
  \label{table:bde-pbe0}
  %%%%%%%%%%%%%%%%%%%%%%%%%%%%
   \begin{adjustwidth}{-\extralength}{0cm}%\centering
 \begin{tabular*}{\fulllength}{@{\extracolsep{\fill}}clcccccc}
\toprule
\textbf{Molecule}   & \textbf{ Method}                         & \textbf{Position }               & \textbf{BDE   } &        \textbf{ IP}   &        \textbf{PDE}     &       \textbf{ PA}    &       \textbf{ ETE  } \\  
\midrule                                                                             
ATV        & UPBE0/IEFPCM                    &  1-OH                   &         93.5   &       107.7  &        23.2    &        24.5  &       106.4  \\
           & UPBE0/IEFPCM                    &  4-OH                   &        109.8   &              &        39.4    &        45.4  &       101.7  \\
\midrule     
ATV        & UB3LYP/IEFPCM                   &  1-OH                   &         91.4   &       107.0  &        22.4    &        23.8  &       105.7  \\
ATV        & UB3LYP/IEFPCM                   &  4-NH                   &         90.2   &              &        21.3    &        44.4  &        83.9  \\
\midrule     
ATV        & ROB3LYP/IEFPCM                  &  1-OH                   &         92.4   &       108.0  &        21.4    &       (23.8) &       106.7  \\      
ATV        & ROB3LYP/IEFPCM                  &  4-NH                   &         92.2   &              &        22.2    &       (44.4) &        85.9  \\  
\midrule     
ATV   & UB3LYP/SMD                 &  1-OH              &     85.9  &   101.2 &    22.7   &    24.0  &  100.0 \\
           & UB3LYP/SMD                 &  4-NH              &     90.7  &         &    27.6   &    44.0  &   84.8 \\
\midrule                                                                    
o-ATV & UB3LYP/IEFPCM              &  5-OH              &     77.5  &   106.9 &     8.7   &    34.4  &   91.8 \\
           & UB3LYP/SMD                 &  5-OH              &     79.6  &   101.0 &    16.8   &    34.6  &   83.1 \\
\midrule                                                            
p-ATV & UB3LYP/IEFPCM              &  5-OH              &     77.4  &   106.2 &     9.2   &    37.9  &   90.8 \\     
           & UB3LYP/SMD                 &  5-OH              &     78.0  &   100.9 &    15.2   &    36.5  &   79.5 \\
\bottomrule
\end{tabular*}
\end{adjustwidth}
  %%%%%%%%%%%%%%%%%%%%%%%%%%%%%%%%%%%%%%%%%%%%%%%%%%%%%%%%%%%%%%%%%%%%%%%%%%%%%%%%%%%%%%

\end{table}

Inspection of Table~\ref{table:bde} and Figure \ref{fig:bde-etc} reveals that the additional 5-OH group
has no notable impact on the O-H bond cleavage at positions 1-OH, 2-OH, and 3-OH, neither homolytic and heterolytic.
BDE for H-atom abstraction at positions 1-OH, 2-OH, and 3-OH in ATV, o-ATV and p-ATV is basically the same.
The differences between the values
calculated by us for ATV, o-ATV, and p-ATV amounting to at most 0.5\,kcal/mol are certainly irrelevant;
recall that we showed recently \cite{Baldea:2022e} that even for much smaller molecules in vacuo DFT/B3LYP calculations with the largest Pople basis set
6-311++G(3df,3pd) are far away from ``chemical'' accuracy ($\sim$1 \,kcal/mol). 
In fact, p-ATV's numerical value of PA = $58.2$\,kcal/mol somewhat differs from ATV's (and o-ATV's) PA~=~$61.5$\,kcal/mol, but if heterolytic
O-H bond cleavage were to occur in p-ATV, it would rather occur at position 1-OH, which has a substantially smaller value PA = $23.8$\,kcal/mol.

With regards to position 4-NH,
the extra (5-)OH-group has a qualitatively different impact on the N-H bond cleavage of o-ATV and p-ATV. 
Notwithstanding
the different values calculated (90.2\,kcal/mol versus 89.3\,kcal/mol), in the above vein
we cannot soberly claim that H-atom abstraction from the NH-group is facilitated by the additional OH-group of o-ATV.
However, the negative impact on the heterolytic N-H bond dissociation is significant. The o-ATV's PA = $49$\,kcal/mol
is larger than the value PA = $44.4$\,kcal/mol calculated for ATV. As of the heterolytic N-H bond dissociation, it is insensitively
affected; the numerical difference between p-ATV's PA = $43.8$\,kcal/mol and ATV's PA = $44.4$\,kcal/mol
obtained within B3LYP/6-31+G(d,p)/IEFPCM is too small to play a role in a sober analysis. Besides, similarly to what we said above,
a heterolytic bond cleavage would occur at the lowest PA's position 1-OH.

The really important effect brought about by the extra OH-group of the hydroxy metabolites is the homolytic bond dissociation at
its position (5-OH). Our calculations demonstrate that this process is substantially less expensive energetically
than H-atom donation from position 1-OH. The calculated BDE values for both o-ATV and p-ATV at this position
are $\sim$77.5\,kcal/mol 
versus the smallest value $\sim$91 \,kcal/mol for ATV at position 1-OH, respectively.
Unlike the extremely similar homolytic bond dissociation, there is a certain difference between o-ATV's and p-ATV's heterolytic
bond dissociation at position 5-OH, as expressed by the PA values (PA=$34.4$\,kcal/mol $\neq$ PA = $37.9$\,kcal/mol, respectively).
However, it is unlikely that this difference in PA's has practical consequences, again because the aforementioned values of PA
are both comfortably larger than the lowest PA = $23.8$\,kcal/mol at position 1-OH, a value that also characterizes the parent ATV~molecule.

In \secname\ref{sec:practice} we will return to the practical implications of the above finding. 
\subsection{Alternative Approaches to the O-H and N-H Bond Strengths: Vibrational Frequencies and Bond Order Indices}
\label{sec:nbo}
Let us start this section with a short digression.
The robustness of a single molecule diode fabricated using the scanning transmission microscopy (STM)
break-junction technique \cite{Tao:03,Tao:03b} can be quantified by the maximum force that the junction
subject to mechanical stretching can withstand. This rupture (pull-off) force $F$ per molecule,
which characterizes the
strength of the chemical bond between electrodes and the terminal (anchoring) atom of the embedded molecule,
can hardly be directly measured. To circumvent this difficulty,
experimentalists use a simple mechanical model which relates $F$
to the vibrational frequency of the pertaining stretching mode $\nu$.
The latter quantity can be easily measured by infrared spectroscopy \cite{Tao:12}.
To exemplify, this is the \ce{Au-S} stretching mode
in benchmark nanojunctions wherein molecules are anchored via thiol groups on gold electrodes.

Transposed to the present context, it is interesting to interrogate the relationship between BDE and
the related stretching frequency. In the same vein, a stronger chemical X-Y bond is intuitively expected
to have not only a larger BDE and a higher stretching frequency $\nu(\ce{X-Y})$ but also a shorter length
and a larger bond order index.

With these in mind, let us examine the correlation of the aforementioned quantities in the presently considered molecules.

\begin{figure}[H]
{
\includegraphics[width=6.8cm]{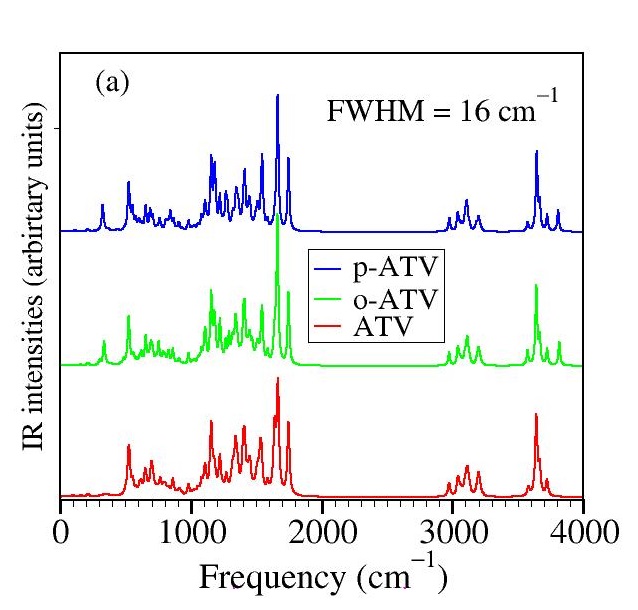}
\includegraphics[width=6.5cm]{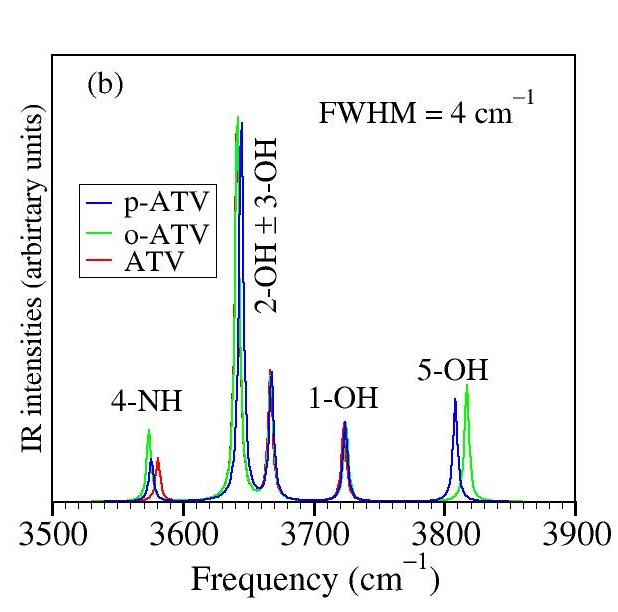}
}
\caption{Infrared spectra calculated for ATV, o-OH-AVT, and p-ATV in methanol using Lorentzian convolution of full width at half maximum (FWHM)
  indicated in the inset:
  (\textbf{a}) in the whole range of frequency and (\textbf{b}) in the range where the O-H and N-H stretching modes are active.
  In all species, stretching modes of 2-OH and 3-OH groups appear as linear and antilinear vibrations rather than separated vibrational modes,
  and this may indicate that a more adequate optimization of the radicals generated by H-atom abstraction at these positions
  (which appear almost degenerate energetically, see pertaining BDE values in {Table}~\ref{table:bde})
  should be done within a multi-reference framework.
  \label{fig:ir}}
\end{figure}

Infrared spectra calculated for ATV, o-OH-AVT, and p-ATV in methanol are depicted in Figure \ref{fig:ir}.
\vspace{-9pt}
\vspace{-6pt}
The behavior visible in Figure \ref{fig:ir}b is surprising for several reasons, e.g.,:

(i) although the BDE of ATV and its metabolites
at position 1-OH is lower than at positions 2-OH and 3-OH,
the streching mode at position 1-OH has a higher frequency than at positions 2-OH and 3-OH;

(ii) although o-ATV and p-ATV have at position 5-OH a smaller BDE than for all OH-positions of the parent ATV,
the 5-OH stretching mode of the metabolites is higher than those of all O-H streching mode of ATV;

(iii) although o-ATV's and ATV's N-H BDE are equal, the frequency of the N-H of the former is smaller than that of the latter;

(iv) although o-ATV's BDE and p-ATV's BDE are different, their N-H streching modes have the same frequency;

(v) although o-ATV and p-ATV have equal BDE at position 5-OH, the o-ATV's O-H streching frequency is higher than
that of p-ATV.

\begin{figure}[H]

\begin{adjustwidth}{-\extralength}{0cm}
\centering

\includegraphics[width=6cm]{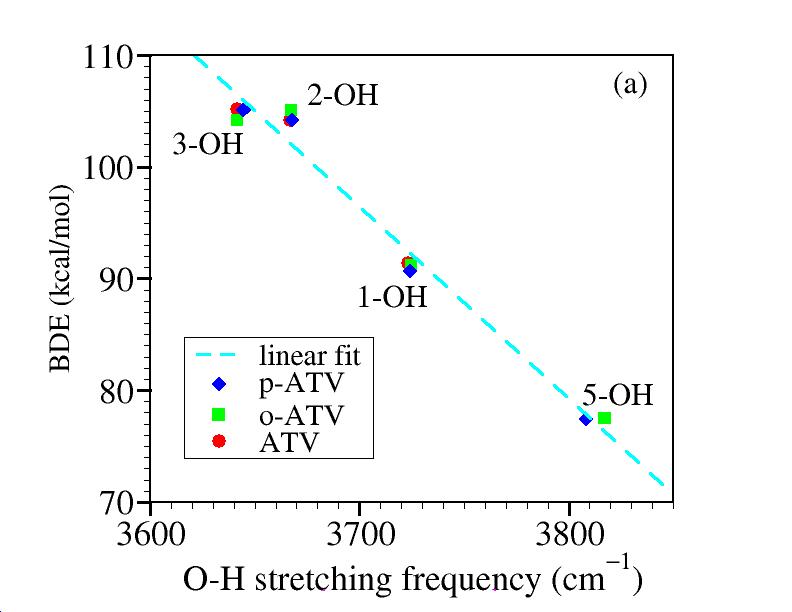}
\includegraphics[width=6cm]{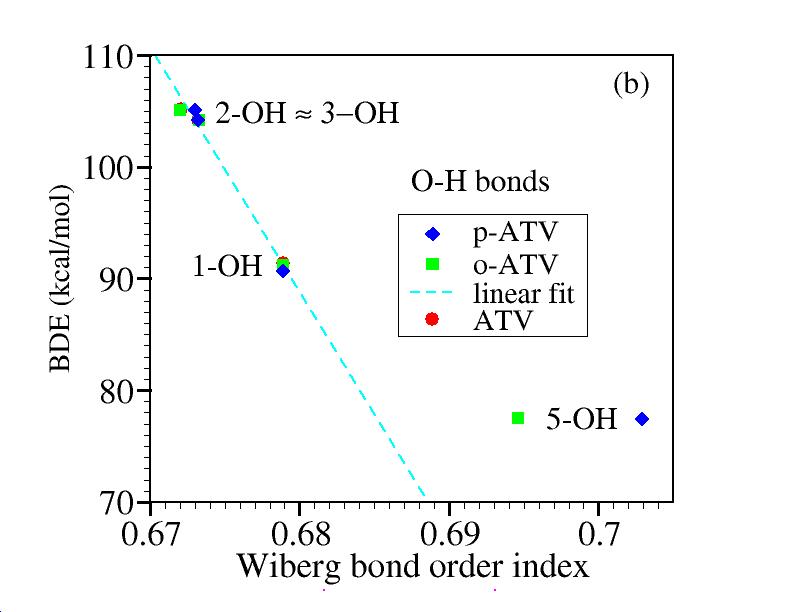}
\includegraphics[width=6cm]{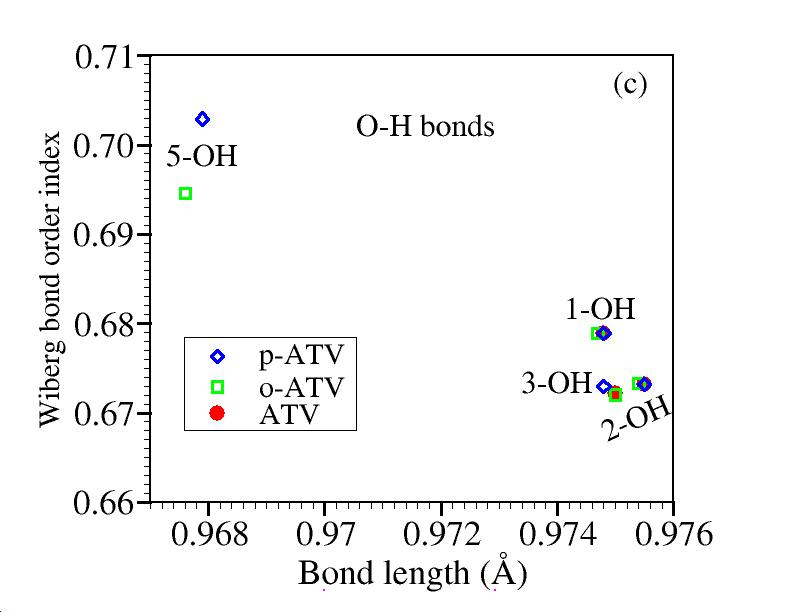}

\end{adjustwidth}
\caption{Results for OH groups of atorvastatin (ATV) and its metabolites o-ATV and p-ATV: (\textbf{a})~bond dissociation energies versus O-H stretching frequencies; 
(\textbf{b}) bond dissociation energies versus Wiberg bond order indices; (\textbf{c}) Wiberg bond order indices versus bond lengths.\label{fig:wbi_oh}}
\end{figure}

\begin{figure}[H]

\begin{adjustwidth}{-\extralength}{0cm}
\centering

\includegraphics[width=6cm]{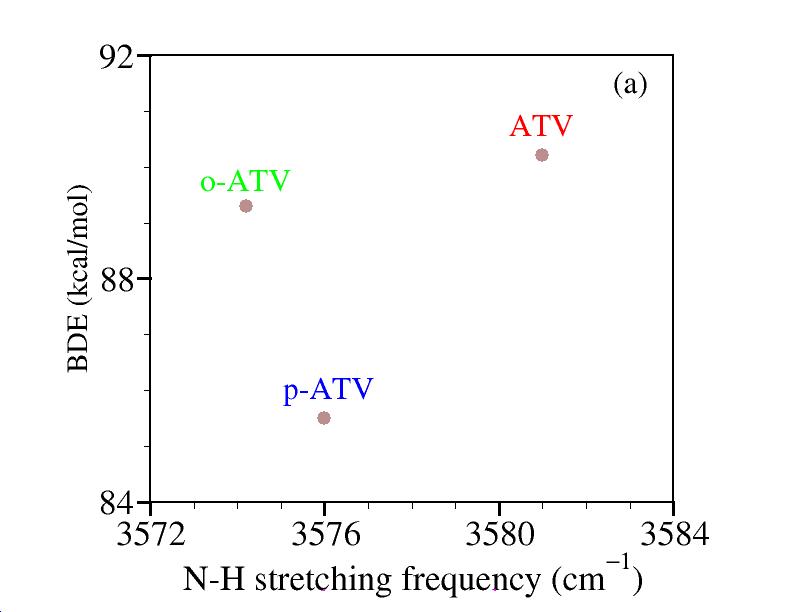}
\includegraphics[width=6cm]{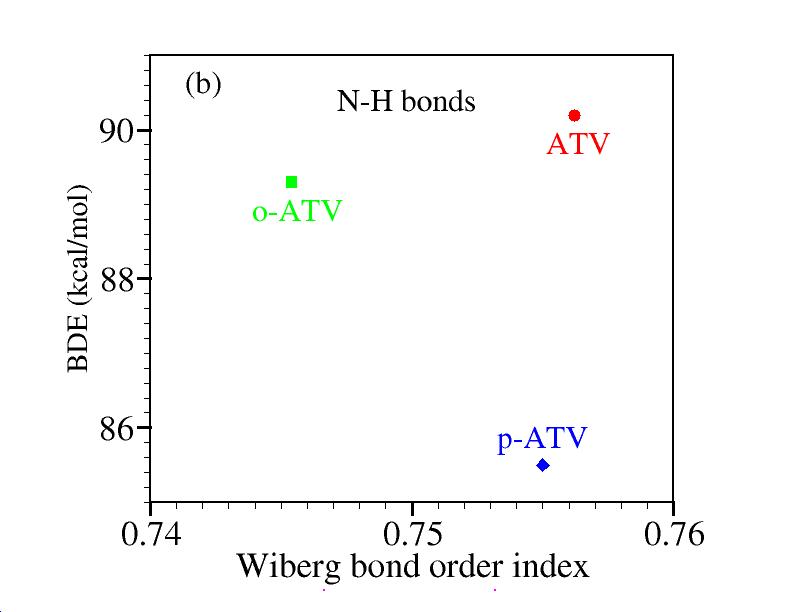}
\includegraphics[width=6cm]{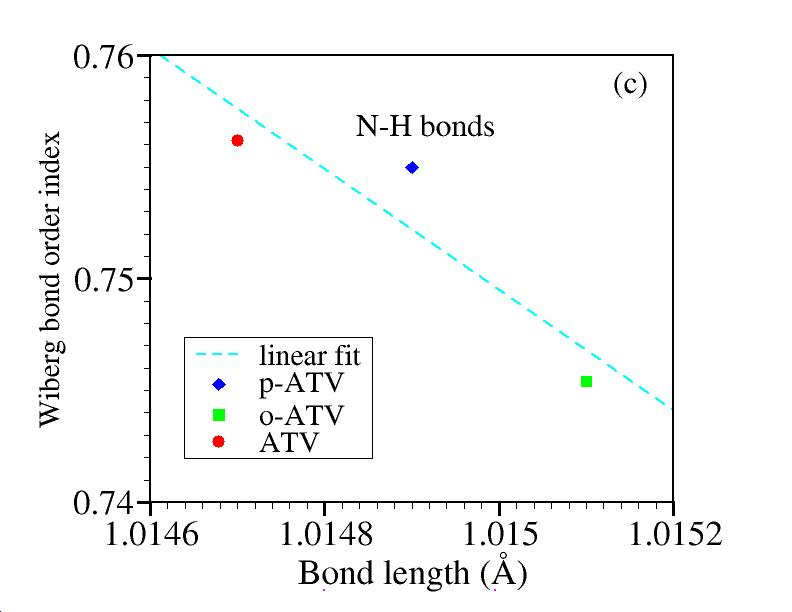}
\end{adjustwidth}
\caption{Results similar to Figure \ref{fig:wbi_oh} but for NH groups:
 (\textbf{a})~bond dissociation energies versus N-H stretching frequencies; 
(\textbf{b}) bond dissociation energies versus Wiberg bond order indices; (\textbf{c}) Wiberg bond order indices versus bond lengths.
  \label{fig:wbi_nh}}
\end{figure}

Counter-intuitive aspects of the relationship BDE versus $\nu$ are visualized in Figures \ref{fig:wbi_oh}a and \ref{fig:wbi_nh}a.

\vspace{-12pt}
%MDPI: there is no exolanation of the subfigure (a--c), please confirm and  add it in the caption.

Let us now switch to bond order indices.
Our results are collected in 
Table~\ref{table:bonds} and Figures \ref{fig:wbi_oh} and \ref{fig:wbi_nh}.

To reiterate, based on straightforward chemical intuition, it would be obvious to expect that stronger chemical bonds
(larger BDE's) possess larger bond order indices. \Figname\ref{fig:wbi_oh}b depicts that for the O-H bonds
of ATV, o-ATV, and p-ATV just the opposite holds true: larger BDE's justly correspond to smaller bond order indices.
As for their N-H bonds, Figure \ref{fig:wbi_oh}b reveals that the dependence is even nonmonotonic.

To avoid misunderstanding, a clarification is in order before ending this analysis. What 
chemical intuition in the above example should not overlook is that a pair of atoms X and Y forming an X-Y chemical bond,
do not merely interact with each other but also with the neighboring atoms in the molecular surrounding.
This is also why a simple (exponential \cite{Pauling:47})
relationship between bond order indices and bond lengths can hold, e.g., for homologous molecular series \cite{Baldea:2019e},
but cannot not hold in general; otherwise one arrives at comparing apples with oranges. \Figsname\ref{fig:wbi_oh} and \ref{fig:wbi_nh} illustrate this again using the
values of Table~\ref{table:bonds}.
BDE values corresponding to different O-H bonds of a given molecule differ from each other
depending on the specific chemical environment. These differences can be visualized by inspecting the spin density landscape
of the various radicals (Figures \ref{fig:atv}, \ref{fig:1-oh-oatv-patv}, \ref{fig:4-nh-oatv-patv}, and \ref{fig:5-oh-oatv-patv}).
The stronger the delocalization in a radical,
the easier is its formation, and the lower is the corresponding BDE value. Inspection of Figures \ref{fig:atv}b and c makes it clear, e.g.,
why ATV's BDE at position 3-OH is higher than that at position 1-OH. 
\begin{table}[H] % [h!]
  \small
  \setlength{\tabcolsep}{6.56mm}
 \caption{Wiberg bond order indices, bond lengths (in {\AA}), vibrational frequencies (in cm$^{-1}$),
      and bond dissociation energies BDE (in kcal/mol) for atorvastatin and its metabolites.}
  \label{table:bonds}

  %%%%%%%%%%%%%%%%%%%%%%%%%%%%
    \begin{tabular*}{\textwidth}{@{\extracolsep{\fill}}lccccc}
      \toprule
   \textbf{Molecule}              & \textbf{Position}   &  \textbf{Wiberg}       & \textbf{ Length}     & \textbf{BDE}            &  \boldmath{$\nu$}    \\  
      \midrule                                                                    
{ATV  }                    &  1-OH      &       0.6789  &  0.9748     &         91.4   &   3723.2  \\
                         &  2-OH      &       0.6732  &  0.9755     &        104.2   &   3667.0  \\
                         &  3-OH      &       0.6721  &  0.9750     &        105.2   &   3641.6  \\
                         &  4-NH      &       0.7562  &  1.0147     &         90.2   &   3581.0  \\

\midrule                                                                                  
{o-ATV}                 &  1-OH      &       0.6789  & 0.9747      &         91.2   &   3724.8  \\
                         &  2-OH      &       0.6733  & 0.9754      &        104.2   &   3641.7  \\
                         &  3-OH      &       0.6720  & 0.9750      &        105.1   &   3667.5  \\
                         &  4-NH      &       0.7454  & 1.0151      &         89.3   &   3574.2  \\
                         &  5-OH      &       0.6946  & 0.9676      &         77.5   &   3817.1  \\
     \midrule                                                                                      
{p-ATV}                 &  1-OH      &       0.6789  & 0.9748      &         90.7   &   3724.2  \\
                         &  2-OH      &       0.6732  & 0.9755      &        104.2   &   3667.9  \\
                         &  3-OH      &       0.6730  & 0.9748      &        105.1   &   3644.7  \\
                         &  4-NH      &       0.7550  & 1.0149      &         85.5   &   3576.0  \\
                         &  5-OH      &       0.7029  & 0.9679      &         77.4   &   3808.4  \\
     \bottomrule                                                                                       
    \end{tabular*}
  %%%%%%%%%%%%%%%%%%%%%%%%%%%%%%%%%%%%%%%%%%%%%%%%%%%%%%%%%%%%%%%%%%%%%%%%%%%%%%%%%%%%%%
   
\end{table}

\subsection{Assessing the Radical Scavenging Activity. A Specific Example}
\label{sec:practice}
Discussion on free radical scavenging and dominant antioxidant mechanism is very often couched by comparing
among themselves values the enthalpies characterizing the HAT, SET-PL, and SPLET of the specific antioxidant(s) under investigation.
Every now and then publications conclude, e.g., that SPLET is the dominant pathway because a certain antioxidant has a ``small'' PA value
or a PA substantially smaller than BDE, or that SET-PL prevails because of the small IP value.
However, it is worth emphasizing that, along with the antioxidant's properties,
a proper evaluation of the antioxidant activity should mandatory consider the specific properties of the radicals
to be eliminated (neutralized).

The small value BDE $\approx \,77.5$\,kcal/mol for o-ATV and p-ATV, substantially smaller than the smallest value (BDE = 90.2\,kcal/mol)
of the parent ATV, is perhaps the most appealing result reported in \secname\ref{sec:bde}.
Still, the ``small'' value mentioned above does not demonstrate \emph{per se} the fact anticipated in Introduction,
namely that o-ATV and p-ATV can scavenge
can scavenge the ubiquitously employed 1,1-diphenyl-2-picrylhydrazyl (DPPH$^\bullet$) radical,
while the parent ATV cannot.

To demonstrate this, one should mandatory consider the pertaining DPPH$^\bullet$ property,
namely the enthalpy release in DPPH$^\bullet$'s neutralization (H-atom affinity)
\begin{equation}
  \ce{DPPH^\bullet} + \ce{H^\bullet} \to \ce{DPPHH} .
\end{equation}

Because it amounts to 80\,kcal/mol \cite{bde-dpphh}, e.g., the reaction
\begin{equation}
  \ce{o-ATV + DPPH^{\bullet}} \to  \ce{o-ATV}5\ce{H^\bullet} + \ce{DPPHH}
  \label{eq-hat-dpph}
\end{equation}
is exothermic. H-atom abstraction from position 5-OH of o-ATV (or p-ATV) costs \linebreak $\sim$77.5\,kcal/mol, a value lower that the enthalpy release of 80\,kcal/mol \cite{bde-dpphh}
in the neutralization of the DPPH$^\bullet$ radical, and this makes the HAT mechanism thermodynamically allowed.
Rephrasing, because the BDE of the \ce{N-H} bond of DPPHH is 80\,kcal/mol \cite{bde-dpphh},
o-ATV (and p-ATV) can scavenge the DPPH$^\bullet$ radical through donating the H-atom at position 5-OH.
On the contrary, the parent ATV cannot. The lowest ATV's BDE (at position 1-OH) amounts to 90.2\,kcal/mol (Table~\ref{table:bde}),
so the HAT pathway is forbidden.

To conclude, we have presented above the first theoretical explanation of the experimental fact \cite{Aviram:98} that
the antioxidant properties of atorvastatin ortho- and para-hydroxy metabolites differ from those of ATV.

By and large, there is a consensus in the literature that HAT is a possible (or even preferred) antioxidant mechanism in the gases phase
but not in polar protic solvents like the presently considered methanol. In this vein, the natural question that arises is:
can o-ATV and p-ATV scavenge the DPPH$^\bullet$ radical in methanol also via SPLET? Can HAT and SPLET coexist? 
While the large IP (Table~\ref{table:bde})
give little chances to an SET-PT pathway, SPLET would a priori be conceivable in view of the ``small'' value of PA,
which is, although not smaller than that of ascorbic acid (as incorrectly \cite{baldea_2022_chemrxiv} claimed
in ref.~\cite{Duque:22})
at least not much larger than the latter (23.8\,kcal/mol for ATV's versus 20.5\,kcal/mol for ascorbic acid,
see ref.~\cite{baldea_2022_chemrxiv}).

In fact, Table~\ref{table:bde} implicitly gives the \emph{negative} answer to this question.
If o-ATV and p-ATV could scavenge DPPH$^\bullet$ via SPLET, then (contrary to experiment \cite{Aviram:98})
the parent ATV could also do the job;
the most favored deprotonation, implying the same enthalpy PA $=$ 23.8\,kcal/mol,
occurs both for ATV and its metabolites at the same 1-OH position, where furthermore the similar 
spin density landscapes (compare Figure \ref{fig:atv}b with Figure \ref{fig:1-oh-oatv-patv})
indicate a similar chemical reactivity.

Still, let us remain in the realm of theory and demonstrate why neither o-ATV nor p-ATV or ATV can scavenge DPPH$^\bullet$ in methanol via SPLET.
To this aim suffice it to consider the first step of SPLET
\begin{equation}
\label{eq-pa-dpph}
\ce{{xATV} + DPPH^{\bullet}} \to \ce{xATV}1\ce{H-} + \ce{DPPHH^{\bullet +}} ,
\end{equation}
where x means ``o-'', ``p-'', or ``nothing''.
Straightforward manipulation allows to express the enthalpy of this reaction as follows
\vspace{-10pt}
\begingroup\makeatletter\def\f@size{9.5}\check@mathfonts
\def\maketag@@@#1{\hbox{\m@th\fontsize{10}{10}\selectfont\normalfont#1}}

\begin{equation}
\label{eq-step1-dpph}
H_r = \underbrace{H(\ce{xATV}1\ce{H}) + H(\ce{H+}) - H(\ce{xATV})}_{\mbox{PA}(\ce{xATV})} -
\underbrace{ H(\ce{DPPH^{\bullet}}) + H(\ce{H+}) - H(\ce{DPPHH^{\bullet +}})}_{\mbox{PDE}(\ce{DPPHH})} .
\end{equation}
\endgroup

Notice that the second brace in \gl~(\ref{eq-step1-dpph}) corresponds to
the proton abstraction from the cation $\ce{DPPHH^{\bullet +}}$ of
the neutralized free radical DPPHH, or alternatively, the PDE pertaining to the 
neutralized free radical DPPHH (cf.~\gl~(\ref{eq-setpt-pde})).

\Gl~(\ref{eq-step1-dpph}) reveals that, to be thermodynamically allowed, the first SPLET step requires 
\begin{equation}
  \label{eq-step1-dpph-2}
  \mbox{PA}(\ce{xATV}) < \mbox{PDE}(\ce{DPPHH}) .
\end{equation}

Our calculations yielded $\mbox{PDE}(\ce{DPPHH}) = 3.9$\,kcal/mol, a value that is not larger 
(as the case if the first SPLET step was allowed) but smaller
than $\mbox{PA}(\ce{xATV}) = 23.8$\,kcal/mol. It now becomes
clear why neither ATV, nor o-ATV or p-ATV can scavenge the DPPH$^\bullet$ radical via SPLET. Their ``small'' PA is not small enough
to fulfill \gl~(\ref{eq-step1-dpph-2}).

\section{Conclusions}
We believe that the present demonstration that atorvastatin ortho- and para-hydroxy metabolites can scavenge the DPPH$^\bullet$
through donating the H-atom at the position of their extra group (5-OH), which is impossible in the parent ATV, is important
not only because it theoretically explains for the first time a behavior revealed in experiment \cite{Aviram:98}
but also because, from a general perspective, it provides further insight into the structure–activity relationship (SAR).

By working out a specific example (\secname\ref{sec:practice})---an analysis that can be straightforwardly
extended to other cases---, we drew attention that an adequate approach to antioxidant's potency should mandatory
account for the thermodynamic properties of the free radicals.
\Gl~(\ref{eq-step1-dpph-2}) expresses a general necessary condition for
thermodynamically allowed SPLET, and its application to specific cases may reveal
that, even in polar solvents, free radical scavenging via this pathway is forbidden not only for ATV-based species. 

In addition, our study emphasize that, while important, e.g., for modeling the temporal evolution of various molecular species
interacting among themselves in a given chemical environment \cite{Baldea:2019e,Baldea:2020b},
the global chemical reactivity indices have no direct relevance for antioxidation. Recall that we saw in \secname\ref{sec:eta}
that quantitative differences of ATV's o-ATV's, and p-ATV's
global chemical reactivity indices are minor. Furthermore, if qualitative differences in these indices were important,
then, contrary to \secsname\ref{sec:bde} and \ref{sec:practice},
o-ATV would have antioxidant properties similar to ATV rather than to p-ATV.

Last but not least, from the perspective of fundamental science, we found (\secname\ref{sec:nbo}) that properties like
bond dissociation enthalpy, bond order index, bond length,
and bond stretching frequency, expected after all to represent alternatives in quantifying the bond strength,
are by no means correlated according to naive intuition.
This finding calls for further quantum chemical efforts aiming at
comprehensively characterizing ATV's, that inherently remained beyond the scope of this study focused on ATV's antioxidant activity.
Finally, the presently reported counter-intutitve relationship between bond stretching frequency and bond strength should 
also be a word of caution for other communities; for example, for the molecular electronics community, wherein
bond stretching frequencies (conveniently obtained via infrared spectroscopy) are used 
to estimate (pull-off) forces that cause the rupture of a junction subject to mechanical stretching \cite{Tao:17}.

\vspace{6pt}

\authorcontributions{Not applicable, this is a single-author paper}%mdpi: please add this part. For research articles with several authors, a short paragraph specifying their individual contributions must be provided. The following statements should be used ``Conceptualization, X.X. and Y.Y.; methodology, X.X.; software, X.X.; validation, X.X., Y.Y. and Z.Z.; formal analysis, X.X.; investigation, X.X.; resources, X.X.; data curation, X.X.; writing---original draft preparation, X.X.; writing---review and editing, X.X.; visualization, X.X.; supervision, X.X.; project administration, X.X.; funding acquisition, Y.Y. All authors have read and agreed to the published version of the manuscript.'', please turn to the  \href{http://img.mdpi.org/data/contributor-role-instruction.pdf}{CRediT taxonomy} for the term explanation. Authorship must be limited to those who have contributed substantially to the work~reported.
\funding{In the initial stage, this research was funded by the German Research Foundation (DFG grant BA 1799/3-2). Computational support from the
state of Baden-W\"urttemberg through bwHPC and the German Research Foundation through
Grant No.~INST 40/575-1 FUGG (bwUniCluster 2.0, bwForCluster/MLS\&WISO 2.0, and JUSTUS 2.0 cluster) is gratefully acknowledged.}

\institutionalreview{Not applicable}%MDPI: please add this part, this is optional. In this section, you should add the Institutional Review Board Statement and approval number, if relevant to your study. You might choose to exclude this statement if the study did not require ethical approval. Please note that the Editorial Office might ask you for further information. Please add “The study was conducted in accordance with the Declaration of Helsinki, and approved by the Institutional Review Board (or Ethics Committee) of NAME OF INSTITUTE (protocol code XXX and date of approval).” for studies involving humans. OR “The animal study protocol was approved by the Institutional Review Board (or Ethics Committee) of NAME OF INSTITUTE (protocol code XXX and date of approval).” for studies involving animals. OR “Ethical review and approval were waived for this study due to REASON (please provide a detailed justification).” OR “Not applicable” for studies not involving humans or animals.

\informedconsent{Not applicable} %MDPI: please add this part, this is optional. Any research article describing a study involving humans should contain this statement. Please add ``Informed consent was obtained from all subjects involved in the study.'' OR ``Patient consent was waived due to REASON (please provide a detailed justification).'' OR ``Not applicable'' for studies not involving humans. You might also choose to exclude this statement if the study did not involve humans. Written informed consent for publication must be obtained from participating patients who can be identified (including by the patients themselves). Please state ``Written informed consent has been obtained from the patient(s) to publish this paper'' if applicable.

\dataavailability{The data that support the findings of this study are available from the author upon reasonable request.
}

\acknowledgments{The author is much indebted to Ederley V\'elez Ortiz for providing valuable details
related to her recent work \cite{Duque:22}. 
}%mdpi: this is as the same as the funding part, please confirm if this can be deleted?

\conflictsofinterest{No conflict of interest to declare} 
\appendixtitles{no} % Leave argument "no" if all appendix headings stay EMPTY (then no dot is printed after "Appendix A"). If the appendix sections contain a heading then change the argument to "yes".
\appendixstart
\appendix
\section[\appendixname~\thesection]{}
\label{app:A}
\begin{table}[H]
\caption{Z$-$matrix of ATV.\label{table:zmat-atv}}
\newcolumntype{C}{>{\centering\arraybackslash}X}
\begin{tabularx}{\textwidth}{CCCCCCCC}
  \toprule
\textbf{Atom}                 &       &                &            &                   &            &                   &             \\
\midrule
    F     &                           &                &            &                   &            &                   &             \\             
    O     &                   1       &           B1   &            &                   &            &                   &             \\ 
    O     &                   1       &           B2   &       2    &              A1   &            &                   &             \\ 
    N     &                   3       &           B3   &       1    &              A2   &       2    &              D1   &       0     \\ 
    C     &                   2       &           B4   &       1    &              A3   &       4    &              D2   &       0     \\ 
    C     &                   3       &           B5   &       1    &              A4   &       4    &              D3   &       0     \\ 
    C     &                   4       &           B6   &       3    &              A5   &       1    &              D4   &       0     \\
\bottomrule
\end{tabularx}
\end{table}

\begin{table}[H]\ContinuedFloat

\caption{\emph{Cont.}
}
\newcolumntype{C}{>{\centering\arraybackslash}X}
\begin{tabularx}{\textwidth}{CCCCCCCC}
  \toprule
\textbf{Atom}                 &       &                &            &                   &            &                   &             \\
\midrule
    H     &                  53       &          B54   &       5    &             A53   &       2    &             D52   &       0     \\ 
   O     &                  50       &          B55   &      47    &             A54   &       4    &             D53   &       0     \\ 
    H     &                  56       &          B56   &      50    &             A55   &      47    &             D54   &       0     \\ 
    O     &                  53       &          B57   &       5    &             A56   &       2    &             D55   &       0     \\ 
    H     &                  58       &          B58   &      53    &             A57   &       5    &             D56   &       0     \\ 
    O     &                   5       &          B59   &       2    &             A58   &       1    &             D57   &       0     \\ 
    H     &                  60       &          B60   &       5    &             A59   &       2    &             D58   &       0     \\
\bottomrule
\end{tabularx}
\end{table}

\vspace{-8pt}
\begin{table}[H]
\caption{Elements of the Z$-$matrix of ATV in methanol optimized as indicated below using 6-31+G/d,p) basis sets.\label{table:zmat-atv-elements}}
\newcolumntype{C}{>{\centering\arraybackslash}X}
\begin{tabularx}{\textwidth}{CCC}
  \toprule
\textbf{Element} &  \textbf{RB3LYP}               &  \textbf{RPBE0}               \\
\midrule
   B1      &          7.63688039	    &         7.57842858 \\ 
   B2      &          9.51376438	    &         9.49797976 \\ 
   B3      &          4.46432846	    &         4.45735736 \\ 
   B4      &          1.22436588	    &         1.22105960 \\ 
   B5      &          1.24094609	    &         1.23460721 \\ 
   B6      &          1.38313941	    &         1.37429179 \\ 
   B7      &          1.39258446	    &         1.38313248 \\ 
   B8      &          4.26421886	    &         4.25026562 \\ 
   B9      &          2.55737607	    &         2.53336720 \\ 
   B10     &          1.39352326	    &         1.38966003 \\ 
   B11     &          1.38854352	    &         1.38531627 \\ 
   B12     &          1.48198028	    &         1.47533742 \\ 
   B13     &          1.47996148	    &         1.47276739 \\ 
   B14     &          2.54735210	    &         2.53021178 \\ 
   B15     &          1.09461863	    &         1.09441258 \\ 
   B16     &          1.09443324	    &         1.09451231 \\ 
   B17     &          1.09548377	    &         1.09521443 \\ 
   B18     &          1.40411033	    &         1.39956240 \\ 
   B19     &          1.08656542	    &         1.08702976 \\ 
   B20     &          1.39590982	    &         1.39231014 \\ 
   B21     &          1.08596263	    &         1.08625533 \\ 
   B22     &          1.39827379	    &         1.39423877 \\ 
   B23     &          1.08557774	    &         1.08587383 \\ 
  B24     &          1.39829551	    &         1.39470047 \\ 
   B25     &          1.08613361	    &         1.08644255 \\ 
   B26     &          1.39684075	    &         1.39270470 \\ 
   B27     &          1.08266424	    &         1.08414373 \\ 
   B28     &          1.40754480	    &         1.40342299 \\ 
  
   B29     &          1.08542085	    &         1.08622867 \\ 
\bottomrule
\end{tabularx}
\end{table}

\begin{table}[H]\ContinuedFloat

\caption{\emph{Cont.}
}
\newcolumntype{C}{>{\centering\arraybackslash}X}
\begin{tabularx}{\textwidth}{CCC}
  \toprule
\textbf{Element} &      \textbf{RB3LYP}      &           \textbf{RPBE0}       \\
\midrule

   B30     &          1.39703298	    &         1.39303810 \\ 
   B31     &          1.08651573	    &         1.08675442 \\ 
   B32     &          1.39866698	    &         1.39488556 \\ 
  B33     &          1.08611015	    &         1.08634947 \\ 
   B34     &          1.39859436	    &         1.39482202 \\ 
   B35     &          1.08646176	    &         1.08669302 \\ 
   B36     &          1.39688653	    &         1.39286539 \\
\bottomrule
\end{tabularx}
\end{table}

\begin{table}[H]\ContinuedFloat

\caption{\emph{Cont.}
}
\newcolumntype{C}{>{\centering\arraybackslash}X}
\begin{tabularx}{\textwidth}{CCC}
  \toprule
\textbf{Element} &      \textbf{RB3LYP}      &           \textbf{RPBE0}       \\
\midrule

   A13     &     106.44925702	    &       106.31547335    \\ 
   A14     &      89.06053861	    &        89.10098490    \\ 
   A15     &      97.89435637	    &        97.80052634    \\ 
   A16     &     142.49393445	    &       142.62052327    \\ 
   A17     &     137.34771594	    &       137.15532746    \\ 
   A18     &     119.50192439	    &       119.48359463    \\ 
   A19     &     120.38662171	    &       120.33363957    \\
\bottomrule
\end{tabularx}
\end{table}

\vspace{-8pt}

\begin{table}[H]\ContinuedFloat

\caption{\emph{Cont.}
}
\newcolumntype{C}{>{\centering\arraybackslash}X}
\begin{tabularx}{\textwidth}{CCC}
  \toprule
\textbf{Element} &         \textbf{RB3LYP}           &         \textbf{ RPBE0}          \\
\midrule  
  A70     &          109.76379473	    &       109.82714850      \\                               
    A71     &           35.55295345	    &        35.64055684      \\ 
   A72     &          108.78209604	    &       109.15267108      \\ 
   A73     &           35.76405946	    &        35.85622083      \\ 
   A74     &          104.88212372	    &       105.29541039      \\ 
   D1      &           23.44707166	    &        22.86899643      \\ 
   D2      &         $-$143.01621213        &      $-$145.69511584	\\
\bottomrule
\end{tabularx}
\end{table}

\begin{table}[H]\ContinuedFloat

\caption{\emph{Cont.}
}
\newcolumntype{C}{>{\centering\arraybackslash}X}
\begin{tabularx}{\textwidth}{CCC}
  \toprule
\textbf{Element} &     \textbf{RB3LYP }     &           \textbf{RPBE0 }           \\
\midrule
    D53     &      135.69337038	   &       135.76291529	        \\ 
   D54     &      154.19232954	   &       153.37488771	        \\ 
   D55     &       $-$1.14110681   &        $-$0.79096165       \\ 
   D56     &        4.84906073	   &         4.80879280	        \\ 
   D57     &      $-$83.86223058   &       $-$85.19930682       \\ 
   D58     &       $-$1.48047919   &        $-$1.48833544       \\ 
   D59     &      161.82475625	   &       161.81654714	        \\ 
   D60     &        8.46484636	   &         8.87965902	        \\ 
   D61     &      147.77400083	   &       149.20907105	        \\ 
   D62     &     $-$178.10763280   &      $-$178.26760363       \\ 
   D63     &       66.16427504	   &        65.99502020	        \\ 
   D64     &      $-$71.70727202   &       $-$71.00674962       \\ 
   D65     &       50.12782255	   &        50.90167682	        \\ 
   D66     &      168.95273147	   &       169.66688376	        \\ 
   D67     &     $-$126.11657682   &      $-$124.36123085       \\ 
   D68     &       54.49143257	   &        54.30008219	        \\ 
   D69     &      $-$62.36290369   &       $-$62.63339700       \\ 
   D70     &      $-$66.30729976   &       $-$66.82133358       \\ 
   D71     &     $-$118.50510691   &      $-$118.64472010       \\ 
   D72     &     $-$128.64978340   &      $-$128.59930105       \\ 
   D73     &     $-$115.36964313   &      $-$115.74853701       \\ 
\bottomrule                                                                                                        
\end{tabularx}                                                                                                     
\end{table}                                                                                                        
         
         \vspace{-8pt}                                                                                                          
\begin{table}[H]
\caption{Z$-$matrix of ATV1H.\label{table:zmat-atv1h}}
\newcolumntype{C}{>{\centering\arraybackslash}X}
\begin{tabularx}{\textwidth}{CCCCCCCC}
  \toprule
\textbf{Atom}                 &       &                &            &                   &            &                   &             \\
\midrule
 F     &                  &           &          &          &      &              &             \\                   
 O     &             1    &        B1 &          &          &      &              &             \\ 
 O     &             1    &        B2 &   2      &      A1  &      &              &             \\ 
 N     &             3    &        B3 &   1      &      A2  &  2   &         D1   &       0     \\  
 C     &             2    &        B4 &   1      &      A3  &  4   &         D2   &       0     \\  
 C     &             3    &        B5 &   1      &      A4  &  4   &         D3   &       0     \\  
 C     &             4    &        B6 &   3      &      A5  &  1   &         D4   &       0     \\
\bottomrule
\end{tabularx}
\end{table}

\begin{table}[H]\ContinuedFloat

\caption{\emph{Cont.}
}
\newcolumntype{C}{>{\centering\arraybackslash}X}
\begin{tabularx}{\textwidth}{CCCCCCCC}
  \toprule
\textbf{Atom}                 &       &                &            &                   &            &                   &             \\
\midrule

 C     &            29    &       B30 &  14      &     A29  & 12   &        D28   &       0     \\  
 H     &            31    &       B31 &  29      &     A30  & 14   &        D29   &       0     \\  
 C     &            31    &       B32 &  29      &     A31  & 14   &        D30   &       0     \\  
 H     &            33    &       B33 &  31      &     A32  & 29   &        D31   &       0     \\  
 C     &            33    &       B34 &  31      &     A33  & 29   &        D32   &       0     \\  
 H     &            35    &       B35 &  33      &     A34  & 31   &        D33   &       0     \\  
 C     &            35    &       B36 &  33      &     A35  & 31   &        D34   &       0     \\
\bottomrule
\end{tabularx}
\end{table}
\vspace{-8pt}
\begin{table}[H]
\caption{Elements of the Z$-$matrix of ATV1H in methanol optimized as indicated below using 6-31+G/d,p) basis sets.\label{table:zmat-atv1h-elements}}
\newcolumntype{C}{>{\centering\arraybackslash}X}
\begin{tabularx}{\textwidth}{CCCC}
  \toprule
\textbf{Element} &                    \textbf{UB3LYP}      &          \textbf{ROB3LYP}             &  \textbf{UPBE0}   \\
\midrule
   B1           &              8.44235812	  &             8.44847690	 &             8.35899747	   \\ 
   B2           &              9.48151258	  &             9.48004054	 &             9.44104389	   \\ 
   B3           &              4.45293368	  &             4.45205992	 &             4.45170279	   \\ 
   B4           &              1.28075664	  &             1.28075532	 &             1.27612637	   \\ 
   B5           &              1.23516216	  &             1.23520893	 &             1.22884511	   \\ 
   B6           &              1.39121019	  &             1.39036552	 &             1.38136414	   \\

\bottomrule
\end{tabularx}
\end{table}

\begin{table}[H]\ContinuedFloat

\caption{\emph{Cont.}
}
\newcolumntype{C}{>{\centering\arraybackslash}X}
\begin{tabularx}{\textwidth}{CCCC}
  \toprule
\textbf{Element} &     \textbf{ UB3LYP }               &      \textbf{ ROB3LY}                 &          \textbf{UPBE0}             \\
\midrule
 B7           &              1.36865245	  &             1.36927061	 &             1.36158073	   \\ 
 B8           &              4.22903226	  &             4.22867569	 &             4.21410980	   \\ 
   B9           &              2.54297346	  &             2.54296359	 &             2.51827392	   \\ 
 B10          &              1.41411564	  &             1.41273105	 &             1.41276955	   \\ 
   B11          &              1.39814133	  &             1.39906265	 &             1.39123927	   \\ 
   B12          &              1.45764082	  &             1.45731631	 &             1.45172243	   \\ 
 B13          &              1.46214878	  &             1.46225268	 &             1.45692318	   \\
\bottomrule
\end{tabularx}
\end{table}

\begin{table}[H]\ContinuedFloat

\caption{\emph{Cont.}
}
\newcolumntype{C}{>{\centering\arraybackslash}X}
\begin{tabularx}{\textwidth}{CCCC}
  \toprule
\textbf{Element} &          \textbf{UB3LYP}                &       \textbf{ROB3LY}                 &          \textbf{UPBE0}         \\
\midrule
B65          &      1.09151271	          &     1.09148630	         &     1.09246030	   \\ 
 B66          &      1.09323237	          &     1.09319168	         &     1.09361961	   \\ 
   B67          &      1.09407784	          &     1.09408957	         &     1.09401377	   \\ 
   B68          &      1.53535621	          &     1.53533608	         &     1.52584909	   \\ 
   B69          &      1.09518094	          &     1.09518217	         &     1.09596063	   \\ 
   B70          &      1.09496161	          &     1.09497326	         &     1.09558744	   \\ 
   B71          &      1.43537977	          &     1.43540753	         &     1.42193713	   \\
\bottomrule
\end{tabularx}
\end{table}

\begin{table}[H]\ContinuedFloat

\caption{\emph{Cont.}
}
\newcolumntype{C}{>{\centering\arraybackslash}X}
\begin{tabularx}{\textwidth}{CCCC}
  \toprule
\textbf{Element} &         \textbf{UB3LYP}                &       \textbf{ROB3LY }                &          \textbf{UPBE0}              \\
\midrule
A48          &      157.42614280	          &   157.44924436	         &   157.60290007	\\ 
   A49          &       94.85343455	          &    94.84153771	         &    95.55489296	\\ 
   A50          &       64.95144615	          &    64.95648523	         &    64.45763585	\\ 
   A51          &      116.19161990	          &   116.19053946	         &   115.94779395	\\ 
   A52          &      106.86069836	          &   106.86047108	         &   106.91121971	\\ 
   A53          &      109.53392738	          &   109.53458463	         &   109.64753221	\\ 
   A54          &       62.07557911	          &    62.08044425	         &    61.96848271	\\
\bottomrule
\end{tabularx}
\end{table}
                  
\begin{table}[H]\ContinuedFloat

\caption{\emph{Cont.}
}
\newcolumntype{C}{>{\centering\arraybackslash}X}
\begin{tabularx}{\textwidth}{CCCC}
  \toprule
\textbf{Element} &         \textbf{UB3LYP}                &       \textbf{ROB3LY }                &          \textbf{UPBE0}              \\
\midrule

   D32          &     $-$0.68679275	          &    $-$0.67915786	         &    $-$0.68858323	       \\ 
   D33          &    179.67314075	          &   179.67322011	         &   179.71709289	       \\ 
   D34          &     $-$0.20109695	          &    $-$0.20087651	         &    $-$0.18063234	       \\ 
   D35          &    179.30402096	          &   179.29000187	         &   179.32243696	       \\ 
   D36          &   $-$130.28427078	          &  $-$130.51938776	         &  $-$130.58633908	       \\ 
   D37          &      1.84741835	          &     1.91470640	         &     1.67453246	       \\ 
   D38          &    179.74200519	          &   179.74365748	         &   179.60779307	       \\
\bottomrule
\end{tabularx}
\end{table}

\begin{adjustwidth}{-\extralength}{0cm}

\reftitle{References}

\end{adjustwidth}

\begin{thebibliography}{999}

\bibitem[Roth(2002)]{Roth:02}
Roth, B.D.
\newblock The Discovery and Development of Atorvastatin, A Potent Novel
  Hypolipidemic Agent.  {\em Prog. Med. Chem.}  \textbf{ 2002}, \emph{40},  1--22.
\newblock
  {{https://doi.org/10.1016/S0079-6468(08)70080-8}}.%mdpi: deleted the extra information ``Elsevier'', plaese confirm.

\bibitem[Mikulic(2021)]{atv-statista}
Mikulic, M.
\newblock Worldwide revenue of Pfizer's Lipitor from 2003 to 2019. {2021}.
\newblock Available online: \url{https://www.statista.com/statistics/254341/pfizers-worldwide-viagra-revenues-since-2003/} (accessed on 11 July 2022).

\bibitem[Alnajjar \em{et~al.}(2021)Alnajjar, Mohamed, and Kawafi]{Alnajjar:21}
Alnajjar, R.; Mohamed, N.; Kawafi, N.
\newblock Bicyclo[1.1.1]Pentane as Phenyl Substituent in Atorvastatin Drug to
  improve Physicochemical Properties: Drug-likeness, DFT, Pharmacokinetics,
  Docking, and Molecular Dynamic Simulation.
\newblock {\em J. Mol. Struct.} {\bf 2021}, {\em 1230},~129628.
\newblock
  {{https://doi.org//10.1016/j.molstruc.2020.129628}}.

\bibitem[Hoffmann and Nowosielski(2008)]{Hoffmann:08}
Hoffmann, M.; Nowosielski, M.
\newblock DFT study on hydroxy acid-lactone interconversion of statins: The
  case of atorvastatin.
\newblock {\em Org. Biomol. Chem.} {\bf 2008}, {\em 6},~3527--3531.
\newblock {{https://doi.org/10.1039/B803342K}}.

\bibitem[Duque \em{et~al.}(2022)Duque, Guerrero, Colorado, Restrepo, and
  Velez]{Duque:22}
Duque, L.; Guerrero, G.; Colorado, J.H.; Restrepo, J.A.; Velez, E.
\newblock Theoretical Insight into mechanism of antioxidant capacity of
  atorvastatin and its o-hydroxy and p-hydroxy metabolites, using DFT methods.
\newblock {\em Comput. Theor. Chem.} {\bf 2022}, \emph{1214},  113758.
\newblock {{https://doi.org//10.1016/j.comptc.2022.113758}}.

\bibitem[B\^aldea(2022)]{baldea_2022_chemrxiv}
B\^aldea, I.
\newblock Critical analysis of radical scavenging properties of atorvastatin in
  methanol recently estimated via density functional theory.
\newblock {\em arXiv} {\bf 2022}, arXiv:2206.13990.
\newblock {{https://doi.org/10.26434/chemrxiv-2022-1bvf1}}.

\bibitem[Portes \em{et~al.}(2007)Portes, Gardrat, and Castellan]{Portes:07}
Portes, E.; Gardrat, C.; Castellan, A.
\newblock A comparative study on the antioxidant properties of
  tetrahydrocurcuminoids and curcuminoids.
\newblock {\em Tetrahedron} {\bf 2007}, {\em 63},~9092--9099.
\newblock {{https://doi.org//10.1016/j.tet.2007.06.085}}.

\bibitem[Aviram \em{et~al.}(1998)Aviram, Rosenblat, Bisgaier, and
  Newton]{Aviram:98}
Aviram, M.; Rosenblat, M.; Bisgaier, C.L.; Newton, R.S.
\newblock Atorvastatin and gemfibrozil metabolites, but not the parent drugs,
  are potent antioxidants against lipoprotein oxidation.
\newblock {\em Atherosclerosis} {\bf 1998}, {\em 138},~271--280.
\newblock
  {{https://doi.org//10.1016/S0021-9150(98)00032-X}}.

\bibitem[Frisch \em{et~al.}(2016)Frisch, Trucks, Schlegel, Scuseria, Robb,
  Cheeseman, Scalmani, Barone, Petersson, Nakatsuji, Li, Caricato, Marenich,
  Bloino, Janesko, Gomperts, Mennucci, Hratchian, Ortiz, Izmaylov, Sonnenberg,
  Williams-Young, Ding, Lipparini, Egidi, Goings, Peng, Petrone, Henderson,
  Ranasinghe, Zakrzewski, Gao, Rega, Zheng, Liang, Hada, Ehara, Toyota, Fukuda,
  Hasegawa, Ishida, Nakajima, Honda, Kitao, Nakai, Vreven, Throssell,
  J.~A.~Montgomery, Peralta, Ogliaro, Bearpark, Heyd, Brothers, Kudin,
  Staroverov, Keith, Kobayashi, Normand, Raghavachari, Rendell, Burant,
  Iyengar, Tomasi, Cossi, Millam, Klene, Adamo, Cammi, Ochterski, Martin,
  Morokuma, Farkas, Foresman, and Fox]{g16}
Frisch, M.J.; Trucks, G.W.; Schlegel, H.B.; Scuseria, G.E.; Robb, M.A.;
  Cheeseman, J.R.; Scalmani, G.; Barone, V.; Petersson, G.A.; Nakatsuji, H.;
  et~al.
\newblock \emph{Gaussian 16, Revision B.01}; Gaussian, Inc.: Wallingford, CT, USA,  2016.%MDPI: newly added, please confirm.
% 

\bibitem[bwHPC(2013)]{bwHPC}
{bwHPC.}
\newblock
\newblock bwHPC 2.0 (JUSTUS cluster\_2.0, bwUniCluster\_2.0, MLS\&WISO\_2.0)
  supported by the State of Baden-W\"{u}rttemberg supported by the state of
  Baden-Württemberg and the German Research Foundation (DFG) through grant no
  INST 40/575-1 FUGG,
  \url{ https://wiki.bwhpc.de/e/Category:BwForCluster_JUSTUS_2}
  {accessed on 12 June 2022.}
  %mdpi: please confirm the type of this ref and provide more information or provide the web link and accessed date.
  % Author: fixed
  
\bibitem[Petersson \em{et~al.}(1988)Petersson, Bennett, Tensfeldt, Al-Laham,
  Shirley, and Mantzaris]{Petersson:88}
Petersson, G.A.; Bennett, A.; Tensfeldt, T.G.; Al-Laham, M.A.; Shirley, W.A.;
  Mantzaris, J.
\newblock A Complete Basis Set Model Chemistry. I. The Total Energies of
  Closed-Shell Atoms and Hydrides of the First-Row Elements.
\newblock {\em J. Chem. Phys.} {\bf 1988}, {\em 89},~2193--2218.
\newblock {{https://doi.org/10.1063/1.455064}}.

\bibitem[Petersson and Al-Laham(1991)]{Petersson:91}
Petersson, G.A.; Al-Laham, M.A.
\newblock A Complete Basis Set Model Chemistry. II. Open-Shell Systems and the
  Total Energies of the First-Row Atoms.
\newblock {\em J. Chem. Phys.} {\bf 1991}, {\em 94},~6081--6090.
\newblock {{https://doi.org/10.1063/1.460447}}.

\bibitem[Lee \em{et~al.}(1988)Lee, Yang, and Parr]{Parr:88}
Lee, C.; Yang, W.; Parr, R.G.
\newblock Development of the Colle-Salvetti correlation-energy formula into a
  functional of the electron density.
\newblock {\em Phys. Rev. B} {\bf 1988}, {\em 37},~785--789.
\newblock {{https://doi.org/10.1103/PhysRevB.37.785}}.

\bibitem[Becke(1988)]{Becke:88}
Becke, A.D.
\newblock Density-Functional Exchange-Energy Approximation with Correct
  Asymptotic Behavior.
\newblock {\em Phys. Rev. A} {\bf 1988}, {\em 38},~3098--3100.
\newblock {{https://doi.org/10.1103/PhysRevA.38.3098}}.

\bibitem[Becke(1993)]{Becke:93a}
Becke, A.D.
\newblock A New Mixing of Hartree-Fock and Local Density-Functional Theories.
\newblock {\em J. Chem. Phys.} {\bf 1993}, {\em 98},~1372--1377.
\newblock {{https://doi.org/10.1063/1.464304}}.

\bibitem[Stephens \em{et~al.}(1994)Stephens, Devlin, Chabalowski, and
  Frisch]{Frisch:94}
Stephens, P.J.; Devlin, J.F.; Chabalowski, C.F.; Frisch, M.J.
\newblock Ab Initio Calculation of Vibrational Absorption and Circular
  Dichroism Spectra Using Density Functional Force Fields.
\newblock {\em J. Phys. Chem.} {\bf 1994}, {\em 98},~11623--11627.
\newblock {{https://doi.org/10.1021/j100096a001}}.

\bibitem[Adamo and Barone(1999)]{Adamo:99}
Adamo, C.; Barone, V.
\newblock Toward Reliable Density Functional Methods without Adjustable
  Parameters: The PBE0 Model.
\newblock {\em J. Chem. Phys.} {\bf 1999}, {\em 110},~6158--6170.
\newblock {{https://doi.org/10.1063/1.478522}}.

\bibitem[Zhao and Truhlar(2006)]{Truhlar:06}
Zhao, Y.; Truhlar, D.G.
\newblock Density Functional for Spectroscopy: No Long-Range Self-Interaction
  Error, Good Performance for Rydberg and Charge-Transfer States, and Better
  Performance on Average than B3LYP for Ground States.
\newblock {\em J. Phys. Chem. A} {\bf 2006}, {\em 110},~13126--13130.
\newblock  {{https://doi.org/10.1021/jp066479k}}.

\bibitem[Zhao and Truhlar(2008)]{Truhlar:08}
Zhao, Y.; Truhlar, D.G.
\newblock The M06 Suite of Density Functionals for Main Group Thermochemistry,
  Thermochemical Kinetics, Noncovalent Interactions, Excited States, and
  Transition Elements: Two New Functionals and Systematic Testing of Four
  M06-Class Functionals and 12 Other Functionals.
\newblock {\em Theor. Chem. Acc.} {\bf 2008}, {\em 120},~215--241.
\newblock {{https://doi.org/10.1007/s00214-007-0310-x}}.

\bibitem[B\^aldea(2022)]{Baldea:2022e}
B\^aldea, I.
\newblock Comprehensive Quantum Chemical Characterization of the
  Astrochemically Relevant \ce{HC_nH} Chain Family. An Attempt to Aid
  Astronomical Observations.
\newblock {\em Adv. Theor. Simul.} {\bf 2022}, 2200244. 
\newblock {{https://doi.org/10.1002/adts.202200244}}.

\bibitem[Kaiser \em{et~al.}(2010)Kaiser, Sun, Lin, Chang, Mebel, Kostko, and
  Ahmed]{Kaiser:10}
Kaiser, R.I.; Sun, B.J.; Lin, H.M.; Chang, A.H.H.; Mebel, A.M.; Kostko, O.;
  Ahmed, M.
\newblock An Experimental and Theoretical Study on the Ionization Energies of
  Polyynes \ce{H\bond{1}(C\bond{3}C)_n\bond{1}H}; n = 1--9).
\newblock {\em Astrophys. J.} {\bf 2010}, {\em 719},~1884.

\bibitem[Tomasi \em{et~al.}(2005)Tomasi, Mennucci, and Cammi]{Tomasi:05}
Tomasi, J.; Mennucci, B.; Cammi, R.
\newblock Quantum Mechanical Continuum Solvation Models.
\newblock {\em Chem. Rev.} {\bf 2005}, {\em 105},~2999--3094.
\newblock PMID: 16092826, {{https://doi.org/10.1021/cr9904009}}.

\bibitem[Canc\`es \em{et~al.}(1997)Canc\`es, Mennucci, and Tomasi]{Cances:97}
Canc\`es, E.; Mennucci, B.; Tomasi, J.
\newblock A new integral equation formalism for the polarizable continuum
  model: Theoretical background and applications to isotropic and anisotropic
  dielectrics.
\newblock {\em J. Chem. Phys.} {\bf 1997}, {\em 107},~3032--3041.
\newblock {{https://doi.org/10.1063/1.474659}}.

\bibitem[Cramer and Truhlar(2008)]{Truhlar:08a}
Cramer, C.J.; Truhlar, D.G.
\newblock A Universal Approach to Solvation Modeling.
\newblock {\em Acc. Chem. Res.} {\bf 2008}, {\em 41},~760--768.
\newblock {{https://doi.org/10.1021/ar800019z}}.

\bibitem[Marenich \em{et~al.}(2008)Marenich, Cramer, and Truhlar]{Truhlar:08b}
Marenich, A.V.; Cramer, C.J.; Truhlar, D.G.
\newblock Perspective on Foundations of Solvation Modeling: The Electrostatic
  Contribution to the Free Energy of Solvation.
\newblock {\em J. Chem. Theory Comput.} {\bf 2008}, {\em 4},~877--887.
\newblock {{https://doi.org/10.1021/ct800029c}}.

\bibitem[Marenich \em{et~al.}(2009)Marenich, Cramer, and Truhlar]{Truhlar:09}
Marenich, A.V.; Cramer, C.J.; Truhlar, D.G.
\newblock Universal Solvation Model Based on Solute Electron Density and on a
  Continuum Model of the Solvent Defined by the Bulk Dielectric Constant and
  Atomic Surface Tensions.
\newblock {\em J. Phys. Chem. B} {\bf 2009}, {\em 113},~6378--6396.
\newblock {{https://doi.org/10.1021/jp810292n}}.

\bibitem[Allouche(2011)]{gabedit}
Allouche, A.R.
\newblock Gabedit: A Graphical User Interface For Computational Chemistry
  Softwares.
\newblock {\em J. Comput. Chem.} {\bf 2011}, {\em 32},~174--182.
\newblock {{https://doi.org/10.1002/jcc.21600}}.

\bibitem[Glendening \em{et~al.}(2012)Glendening, Badenhoop, Reed, Carpenter,
  Bohmann, Morales, and Weinhold]{NBO:6.0}
Glendening, E.; Badenhoop, J.; Reed, A.; Carpenter, J.; Bohmann, J.; Morales,
  C.; Weinhold, F.
  \newblock NBO Code Version 6.0.  2012.
\newblock {{https://nbo6.chem.wisc.edu/ (accessed: April 2022).}} 
  %mdpi: please confirm the type of this ref and provide more information or provide the web link and accessed date.

\bibitem[Wiberg(1968)]{Wiberg:68}
Wiberg, K.B.
\newblock Application of the Pople-Santry-Segal CNDO Method to the
  Cyclopropylcarbinyl and Cyclobutyl Cation and to Bicyclobutane.
\newblock {\em Tetrahedron} {\bf 1968}, {\em 24},~1083--1096.
\newblock {{https://doi.org//10.1016/0040-4020(68)88057-3}}.

\bibitem[Mayer(2007)]{Mayer:07}
Mayer, I.
\newblock Bond Order and Valence Indices: A Personal Account.
\newblock {\em J. Comput. Chem.} {\bf 2007}, {\em 28},~204--221.
\newblock {{https://doi.org//10.1002/jcc.20494}}.

\bibitem[B\^aldea(2022)]{baldea_2022b}
B\^aldea, I.
\newblock Chemical bonding in representative astrophysically relevant neutral,
  cation, and anion HCnH chains.
\newblock {\em ChemRxiv} {\bf 2022}.
\newblock {{https://doi.org/10.26434/chemrxiv-2022-h0pzl}}.

\bibitem[Parr and Yang(1989)]{Parr:89}
Parr, R.G.; Yang, W.
\newblock {\em Density-Functional Theory of Atoms and Molecules}; Oxford
  University Press: Clarendon, UK,  1989; 
\newblock p.~149.

\bibitem[G\'azquez \em{et~al.}(2007)G\'azquez, Cedillo, and Vela]{Gazquez:07}
G\'azquez, J.L.; Cedillo, A.; Vela, A.
\newblock Electrodonating and Electroaccepting Powers.
\newblock {\em J. Phys. Chem. A} {\bf 2007}, {\em 111},~1966--1970.
\newblock PMID: 17305319, {{https://doi.org/10.1021/jp065459f}}.

\bibitem[K~Rajan and C.K.K.~Muraleedharan(2018)]{Rajan:18}
K~Rajan, V.; C.K.K.~Muraleedharan, H.
\newblock The natural food colorant Peonidin from cranberries as a potential
  radical scavenger- A DFT based mechanistic analysis.
\newblock {\em Food Chem.} {\bf 2018}, {\em 262}, 184--190.
\newblock {{https://doi.org/10.1016/j.foodchem.2018.04.074}}.

\bibitem[B\^aldea(2019{\natexlab{a}})]{Baldea:2019f}
B\^aldea, I.
\newblock Impact of Molecular Conformation on Transport and Transport-Related
  Properties at the Nanoscale.
\newblock {\em Appl. Surf. Sci.} {\bf 2019}, {\em 487},~593--600.
\newblock {{https://doi.org//10.1016/j.apsusc.2019.05.112}}.

\bibitem[B\^aldea(2019{\natexlab{b}})]{Baldea:2019g}
B\^aldea, I.
\newblock Alternation of Singlet and Triplet States in Carbon-Based Chain
  Molecules and Its Astrochemical Implications: Results of an Extensive
  Theoretical Study.
\newblock {\em Adv. Theory Simul.} {\bf 2019}, {\em 2},~1900084.
\newblock {{https://doi.org/10.1002/adts.201900084}}.

\bibitem[Burke(2012)]{Burke:12}
Burke, K.
\newblock Perspective on density functional theory.
\newblock {\em J. Chem. Phys.} {\bf 2012}, {\em 136},~150901.
\newblock {{https://doi.org/http://dx.doi.org/10.1063/\linebreak 1.4704546}}.

\bibitem[B\^aldea(2014)]{Baldea:2014c}
B\^aldea, I.
\newblock A Quantum Chemical Study from a Molecular Transport Perspective:
  Ionization and Electron Attachment Energies for Species Often Used to
  Fabricate Single-Molecule Junctions.
\newblock {\em Faraday Discuss.} {\bf 2014}, {\em 174},~37--56.
\newblock {{https://doi.org/10.1039/C4FD00101J}}.

\bibitem[Kohn \em{et~al.}(1996)Kohn, Becke, and Parr]{Kohn:96}
Kohn, W.; Becke, A.D.; Parr, R.G.
\newblock Density Functional Theory of Electronic Structure.
\newblock {\em J. Chem. Phys.} {\bf 1996}, {\em 100},~12974--12980.
% 
% 
\newblock {Notice that after eq.~(19), the authors of this reference state: ``The individual
                  eigenfunctions and eigenvalues, $\varphi_j$ and
                  $\epsilon_j$, of the Kohn-Sham equations have no strict
                  physical significance \ldots''}
\newblock 
  {{https://doi.org/10.1021/jp960669l}}.

\bibitem[Godby \em{et~al.}(1988)Godby, Schl\"uter, and Sham]{Godby:88}
Godby, R.W.; Schl\"uter, M.; Sham, L.J.
\newblock Self-energy operators and exchange-correlation potentials in
  semiconductors.
\newblock {\em Phys. Rev. B} {\bf 1988}, {\em 37},~10159--10175.
\newblock {{https://doi.org/10.1103/PhysRevB.37.10159}}.

\bibitem[Fiorentini and Baldereschi(1995)]{Baldereschi:95}
Fiorentini, V.; Baldereschi, A.
\newblock Dielectric Scaling of the Self-Energy Scissor Operator in
  Semiconductors and Insulators.
\newblock {\em Phys. Rev. B} {\bf 1995}, {\em 51},~17196--17198.
\newblock {{https://doi.org/10.1103/PhysRevB.51.17196}}.

\bibitem[B\^aldea(2013)]{Baldea:2013c}
B\^aldea, I.
\newblock Demonstrating Why DFT-Calculations For Molecular Transport in
  Solvents Need Scissor Corrections.
\newblock {\em Electrochem. Commun.} {\bf 2013}, {\em 36},~19--21.
\newblock
  {{https://doi.org/10.1016/j.elecom.2013.08.027}}.

\bibitem[B\^aldea(2020{\natexlab{a}})]{Baldea:2020c}
B\^aldea, I.
\newblock Profiling \ce{C4N} Radicals of Astrophysical Interest.
\newblock {\em Mon. Not. R. Astron. Soc.} {\bf 2020}, {\em 493},~2506--2510. 
\newblock {{https://doi.org/10.1093/\linebreak mnras/staa455}}.

\bibitem[B\^aldea(2020{\natexlab{b}})]{Baldea:2020e}
B\^aldea, I.
\newblock Profiling Astrophysically Relevant \ce{MgC4H} Chains. An Attempt to
  Aid Astronomical Observations.
\newblock {\em Mon. Not. R. Astron. Soc.} {\bf 2020}, {\em 498},~4316--4326.
\newblock {{https://doi.org/10.1093/mnras/staa2354}}.

\bibitem[Zhou and Parr(1999)]{Parr:99}
Zhou, Z.; Parr, R.G.
\newblock Electrophilicity Index.
\newblock {\em J. Am. Chem. Soc.} {\bf 1999}, {\em 121},~1922--1924.

\bibitem[G\'azquez(2008)]{Gazquez:08}
G\'azquez, J.L.
\newblock {Perspectives on the Density Functional Theory of Chemical
  Reactivity}.
\newblock {\em {J. Mex. Chem. Soc.}} {\bf 2008}, {\em
  52},~3--10.

\bibitem[Domingo \em{et~al.}(2002)Domingo, Aurell, P\'erez, and
  Contreras]{Domingo:02}
Domingo, L.R.; Aurell, M.; P\'erez, P.; Contreras, R.
\newblock Quantitative characterization of the global electrophilicity power of
  common diene/dienophile pairs in Diels-Alder reactions.
\newblock {\em Tetrahedron} {\bf 2002}, {\em 58},~4417--4423.
\newblock
  {{\textls[-15]{https://doi.org//10.1016/S0040-4020(02)00410-6}}}.

\bibitem[Burton \em{et~al.}(1985)Burton, Doba, Gabe, Hughes, Lee, Prasad, and
  Ingold]{Burton:85}
Burton, G.W.; Doba, T.; Gabe, E.; Hughes, L.; Lee, F.L.; Prasad, L.; Ingold,
  K.U.
\newblock Autoxidation of biological molecules. 4. Maximizing the antioxidant
  activity of phenols.
\newblock {\em J. Am. Chem. Soc.} {\bf 1985}, {\em 107},~7053--7065.
\newblock {{https://doi.org/10.1021/ja00310a049}}.

\bibitem[de~Heer \em{et~al.}(2000)de~Heer, Mulder, Korth, Ingold, and
  Lusztyk]{deHeer:00}
de~Heer, M.I.; Mulder, P.; Korth, H.G.; Ingold, K.U.; Lusztyk, J.
\newblock Hydrogen Atom Abstraction Kinetics from Intramolecularly Hydrogen
  Bonded Ubiquinol-0 and Other (Poly)methoxy Phenols.
\newblock {\em J. Am. Chem. Soc.} {\bf 2000}, {\em
  122},~2355--2360. 
\newblock {{https://doi.org/10.1021/ja9937674}}.

\bibitem[Mayer and Salvador(2004)]{Mayer:04}
Mayer, I.; Salvador, P.
\newblock Overlap populations, bond orders and valences for ``fuzzy'' atoms.
\newblock {\em Chem. Phys. Lett.} {\bf 2004}, {\em 383},~368 -- 375.
\newblock {{https://doi.org//10.1016/j.cplett.2003.11.048}}.

\bibitem[Jovanovic \em{et~al.}(1994)Jovanovic, Steenken, Tosic, Marjanovic, and
  Simic]{Jovanovic:94}
Jovanovic, S.V.; Steenken, S.; Tosic, M.; Marjanovic, B.; Simic, M.G.
\newblock Flavonoids as Antioxidants.
\newblock {\em J. Am. Chem. Soc.} {\bf 1994}, {\em
  116},~4846--4851. 
\newblock {{https://doi.org/10.1021/ja00090a032}}.

\bibitem[Jovanovic \em{et~al.}(1996)Jovanovic, Steenken, Hara, and
  Simic]{Jovanovic:96}
Jovanovic, S.V.; Steenken, S.; Hara, Y.; Simic, M.G.
\newblock Reduction potentials of flavonoid and model phenoxyl radicals. Which
  ring in flavonoids is responsible for antioxidant activity?
\newblock {\em J. Chem. Soc.  Perkin Trans. 2} {\bf 1996},   2497--2504.
\newblock {{https://doi.org/10.1039/P29960002497}}.

\bibitem[Litwinienko and Ingold(2003)]{Litwinienko:03}
Litwinienko, G.; Ingold, K.U.
\newblock Abnormal Solvent Effects on Hydrogen Atom Abstractions. 1. The
  Reactions of Phenols with 2,2-Diphenyl-1-picrylhydrazyl (DPPH$^\bullet$) in
  Alcohols.
\newblock {\em  J. Org. Chem.} {\bf 2003}, {\em
  68},~3433--3438. 
\newblock PMID: 12713343, {{https://doi.org/10.1021/jo026917t}}.

\bibitem[Litwinienko and Ingold(2004)]{Litwinienko:04}
Litwinienko, G.; Ingold, K.U.
\newblock Abnormal Solvent Effects on Hydrogen Atom Abstraction. 2. Resolution
  of the Curcumin Antioxidant Controversy. The Role of Sequential Proton Loss
  Electron Transfer.
\newblock {\em  J. Org. Chem.} {\bf 2004}, {\em
  69},~5888--5896. 
\newblock {{https://doi.org/10.1021/jo049254j}}.

\bibitem[Jones and Gunnarsson(1989)]{Gunnarson:89}
Jones, R.O.; Gunnarsson, O.
\newblock The Density Functional Formalism, Its Applications and Prospects.
\newblock {\em Rev. Mod. Phys.} {\bf 1989}, {\em 61},~689--746.
\newblock {{https://doi.org/10.1103/RevModPhys.61.689}}.

\bibitem[B\^aldea(2012)]{Baldea:2012i}
B\^aldea, I.
\newblock Extending the Newns-Anderson Model to Allow Nanotransport Studies
  Through Molecules with Floppy Degrees of Freedom.
\newblock {\em Europhys. Lett.} {\bf 2012}, {\em 99},~47002.
\newblock {{https://doi.org/10.1209/0295-5075/99/47002}}.

\bibitem[Fifen(2013)]{Fifen:13}
Fifen, J.J.
\newblock Thermodynamics of the Electron Revisited and Generalized.
\newblock {\em J. Chem. Theory Comput.} {\bf 2013}, {\em
  9},~3165--3169. 
\newblock {{https://doi.org/10.1021/ct400212t}}.

\bibitem[Fifen \em{et~al.}(2014)Fifen, Dhaouadi, and Nsangou]{Fifen:14}
Fifen, J.J.; Dhaouadi, Z.; Nsangou, M.
\newblock Revision of the Thermodynamics of the Proton in Gas Phase.
\newblock {\em  J. Phys. Chem. A} {\bf 2014}, {\em
  118},~11090--11097. 
\newblock  {{https://doi.org/10.1021/jp508968z}}.

\bibitem[Markovic \em{et~al.}(2016)Markovic, Tosovic, Milenkovic, and
  Markovic]{Markovic:16}
Markovic, Z.; Tosovic, J.; Milenkovic, D.; Markovic, S.
\newblock Revisiting the solvation enthalpies and free energies of the proton
  and electron in various solvents.
\newblock {\em Comput. Theor. Chem.} {\bf 2016}, {\em
  1077},~11--17.
\newblock 
  {{https://doi.org//10.1016/j.comptc.2015.09.007}}.

\bibitem[Rimarcik \em{et~al.}(2010)Rimarcik, Lukes, Klein, and
  Ilcin]{Rimarcik:10}
Rimarcik, J.; Lukes, V.; Klein, E.; Ilcin, M.
\newblock Study of the solvent effect on the enthalpies of homolytic and
  heterolytic N-H bond cleavage in p-phenylenediamine and
  tetracyano-p-phenylenediamine.
\newblock {\em J. Mol. Struct. THEOCHEM} {\bf 2010}, {\em
  952},~25--30.
\newblock
  {{https://doi.org//10.1016/j.theochem.2010.04.002}}.

\bibitem[Xu and Tao(2003)]{Tao:03}
Xu, B.; Tao, N.J.
\newblock Measurement of Single-Molecule Resistance by Repeated Formation of
  Molecular Junctions.
\newblock {\em Science} {\bf 2003}, {\em 301},~1221--1223. 
\newblock {{https://doi.org/10.1126/science.1087481}}.

\bibitem[Xu \em{et~al.}(2003)Xu, Xiao, and Tao]{Tao:03b}
Xu, B.; Xiao, X.; Tao, N.J.
\newblock Measurements of Single-Molecule Electromechanical Properties.
\newblock {\em J. Am. Chem. Soc.} {\bf 2003}, {\em 125},~16164--16165. 
\newblock {{https://doi.org/10.1021/ja038949j}}.

\bibitem[Bruot \em{et~al.}(2011)Bruot, Hihath, and Tao]{Tao:12}
Bruot, C.; Hihath, J.; Tao, N.
\newblock Mechanically controlled molecular orbital alignment in single
  molecule junctions.
\newblock {\em Nat. Nano} {\bf 2011}, {\em 7},~35--40.
\newblock {{https://doi.org/10.1038/nnano.2011.212}}.

\bibitem[Pauling(1947)]{Pauling:47}
Pauling, L.
\newblock Atomic Radii and Interatomic Distances in Metals.
\newblock {\em J. Am. Chem. Soc.} {\bf 1947}, {\em 69},~542--553. 
\newblock {{https://doi.org/10.1021/\linebreak ja01195a024}}.

\bibitem[B\^aldea(2019)]{Baldea:2019e}
B\^aldea, I.
\newblock Long Carbon-Based Chains of Interstellar Medium Can Have a Triplet
  Ground State. Why Is This Important for Astrochemistry?
\newblock {\em ACS Earth Space Chem.} {\bf 2019}, {\em 3},~863--872. 
\newblock {{https://doi.org/10.1021/acsearthspacechem.9b00008}}.

\bibitem[Luo(2003)]{bde-dpphh}
Luo, Y.R.  (Ed.) 
\newblock {\em Handbook of Bond Dissociation Energies in Organic Compounds};
  CRC Press:  Boca Raton, FL, USA,  2003; 
\newblock p. 239.  {{https://doi.org/10.1201/9781420039863}}.

\bibitem[B\^aldea(2020)]{Baldea:2020b}
B\^aldea, I.
\newblock Extensive Quantum Chemistry Study of Neutral and Charged \ce{C4N}
  Chains: An Attempt to Aid Astronomical Observations.
\newblock {\em ACS Earth Space Chem.} {\bf 2020}, {\em 4},~434--448.
\newblock {{https://doi.org/10.1021/acsearthspacechem.9b00321}}.

\bibitem[Li \em{et~al.}(2017)Li, Haworth, Xiang, Ciampi, Coote, and
  Tao]{Tao:17}
Li, Y.; Haworth, N.L.; Xiang, L.; Ciampi, S.; Coote, M.L.; Tao, N.
\newblock Mechanical Stretching-Induced Electron-Transfer Reactions and
  Conductance Switching in Single Molecules.
\newblock {\em J. Am. Chem. Soc.} {\bf 2017}, {\em 139},~14699--14706. 
\newblock  {{https://doi.org/10.1021/jacs.7b08239}}.

\end{thebibliography}
\end{document}